\documentclass[prb,twocolumn,amsmath,amssymb,showpacs,floatfix]{revtex4}

\usepackage{graphicx,bm,epsfig}
\usepackage{color}

\makeatletter
\def\graphicscale{\twocolumn@sw{0.33}{0.4}}
\makeatother

\def\spose#1{\hbox to 0pt{#1\hss}}
\def\lesssim{\mathrel{\spose{\lower 3pt\hbox{$\mathchar"218$}}
 \raise 2.0pt\hbox{$\mathchar"13C$}}}
\def\gtrsim{\mathrel{\spose{\lower 3pt\hbox{$\mathchar"218$}}
 \raise 2.0pt\hbox{$\mathchar"13E$}}}

\def\<{\langle}
\def\>{\rangle}

\newcommand*{\beq}{\begin{eqnarray}}
\newcommand*{\eeq}{\end{eqnarray}}
\newcommand*{\bea}{\begin{eqnarray}}
\newcommand*{\eea}{\end{eqnarray}}

\def\simge{\mathrel{%
       \rlap{\raise 0.511ex \hbox{$>$}}{\lower 0.511ex \hbox{$\sim$}}}}
\def\simle{\mathrel{
       \rlap{\raise 0.511ex \hbox{$<$}}{\lower 0.511ex \hbox{$\sim$}}}}

\begin{document}

\title{Consistent coarse-graining strategy for polymer solutions \\
   in the thermal crossover from good to $\theta$ solvent.}

\author{Giuseppe D'Adamo}
\email{giuseppe.dadamo@aquila.infn.it}
\affiliation{Dipartimento di Fisica, Sapienza Universit\`a di Roma,
P.le Aldo Moro 2, I-00185 Roma, Italy}
\author{Andrea Pelissetto}
\email{andrea.pelissetto@roma1.infn.it}
\affiliation{Dipartimento di Fisica, Sapienza Universit\`a di Roma and
INFN, Sezione di Roma I, P.le Aldo Moro 2, I-00185 Roma, Italy}
\author{Carlo Pierleoni}
\email{carlo.pierleoni@aquila.infn.it}
\affiliation{Dipartimento di Scienze Fisiche e Chimiche, Universit\`a dell'Aquila and
CNISM, UdR dell'Aquila, V. Vetoio 10, Loc. Coppito, I-67100  L'Aquila, Italy}

\date{\today}

\begin{abstract}

We extend our previously developed coarse-graining strategy for linear polymers 
with a tunable number $n$
of effective atoms (blobs) per chain 
[D'Adamo {\em et al.}, J. Chem. Phys.  \textbf{137}, 4901 (2012)]
to polymer systems in thermal crossover between the 
good-solvent and the $\theta$ regimes.
We consider the thermal crossover in the 
region in which tricritical effects can be neglected, i.e., not too close
to the $\theta$ point, for a wide range of chain volume fractions
$\Phi=c/c^*$ ($c^*$ is the overlap concentration), up to $\Phi \approx 30$. 
Scaling crossover functions for global properties of the solution are obtained 
by Monte Carlo simulations of the Domb-Joyce model with suitably rescaled 
on-site repulsion. They provide the input data 
to develop a minimal coarse-grained model with four blobs per chain
(tetramer model).
As in the good-solvent case, the coarse-grained model potentials 
are derived at zero density, thus avoiding 
the inconsistencies related to the use of state-dependent potentials. 
We find that the coarse-grained model 
reproduces the properties of the underlying, full-monomer system 
up to some reduced density $\Phi$ 
which increases when lowering the temperature towards the $\theta$ state.
Close to the lower-temperature crossover boundary, the tetramer model is 
accurate at least up to $\Phi\simeq 10$, while near the good-solvent regime 
reasonably accurate results are obtained up to $\Phi\simeq 2$.
The density region in which the coarse-grained model is predictive can be 
enlarged by developing coarse-grained models with more blobs per chain. 
We extend the strategy used in the good-solvent case to the 
crossover regime. This requires a proper treatment of the length rescalings
as before, but also
a proper temperature redefinition as the number of blobs is increased.
The case $n=10$ is investigated in detail. We obtain the potentials for such
finer-grained model starting from the tetramer ones.
Comparison with full-monomer results 
shows that the density region in which accurate predictions can be obtained is
significantly wider than that corresponding to the tetramer case.
\end{abstract}


\maketitle

\section{Introduction}
The phase diagram of solutions of linear polymers 
and their large-scale structure 
(i.e., on the scale of the coil size) are well understood and can be explained 
by invoking the physical picture of polymers in implicit solvent with 
effective (solvent-mediated) interactions among the monomers. 
\cite{Flory,deGennes-79,Freed-87,dCJ-book,Schaefer-99} 
When the number $L$ of monomers per chain (degree of polymerization) is large,
such chains exhibit scale invariance, meaning that, in suitably rescaled 
units of density, temperature, and chain length, solutions of chemically 
distinct polymers and solvents obey universal scaling relations,
i.e., global properties of the solution are characterized by 
{\em universal} exponents and scaling functions. 
Central properties are the single-chain radius of gyration $R_g$ and the 
equation of state for the osmotic pressure $\Pi$. For small 
concentrations, $\Pi$ admits the expansion\cite{footPsi}
\begin{equation} 
Z=\frac{\beta \Pi}{c} = 1 + B_2 c + B_3 c^2+ o(c^3), 
\label{eq:virial}
\end{equation}
where $\beta=1/(k_B T)$, $B_k$ are the usual  virial coefficients, and 
$c=N/V$ is the polymer density ($N$ chains in a volume $V$). The coefficients
$B_k$ depend on all chemical details. On the other hand, 
the dimensionless ratios $A_k=B_k\hat{R}_g^{-3(k-1)}$ ($\hat{R}_g$ 
is the zero-density radius of gyration) have a universal
limit, i.e., they are independent of chemical details, for large values 
of $L$. 

Scaling functions and exponents are influenced by the nature of 
the interactions among 
the monomers. Three regimes can be distinguished.\cite{DaoudJannink}
i) The {\em good-solvent} regime, usually observed at high enough temperature, 
in which interactions are dominated by the pairwise repulsion and the 
single-chain size grows\cite{Flory,deGennes-72} as $L^\nu$, where 
\cite{Clisby} $\nu = 0.587597(7)\simeq 3/5$ is 
the Flory exponent. The second virial coefficient is positive and \cite{CMP-06}
$A_2 = 5.500(3)$ for $L\to \infty$.
ii) The {\em $\theta$-regime} at lower temperature, in which the repulsive and 
attractive contributions in the two-body and three-body effective interactions 
are vanishingly small and of the same order of magnitude. In the 
scaling limit $L\to\infty$, this regime collapses to a single 
temperature $T_{\theta}$, called  $\theta$ temperature. At $T=T_\theta$ 
chains are Gaussian (ideal) with scaling exponent $\nu=1/2$ and 
the second virial combination $A_2$ vanishes. 
However, for linear chains of finite length, 
as is the case in experiments and computer simulations, 
this regime is observed in a temperature interval around $T_\theta$ of the 
order of $1/\sqrt{L}$. In analogy with the dilute-gas case,
one often defines the ``Boyle" temperature $T_B$, as the value of $T$ at 
which the second virial coefficient $B_2$ vanishes.
$T_B$ is generally larger then $T_{\theta}$, converges to 
$T_\theta$ as $L\to \infty$ [$T_B = T_{\theta}+O(L^{-1/2}(\ln L)^{-7/11})$], 
and is sometimes used to estimate $T_{\theta}$ numerically.\cite{Grassberger1} 
iii) At even lower temperatures the two-body effective interactions become 
predominantly attractive (negative second virial coefficient), the chains are 
segregated by the solvent and below a critical temperature $T_c$ the solution 
separates in a chain-poor solution where chains are in the collapsed state, 
and a solvent-poor solution where the chain statistics approaches the 
one in the melt state (ideal at large scales). Again,
 in the scaling limit $L\to\infty$ the critical temperature $T_c$ 
tends to $T_{\theta}$.
\cite{deGennes-79,Freed-87,dCJ-book}

\begin{figure}[t]
\begin{center}
\begin{tabular}{c}
\epsfig{file=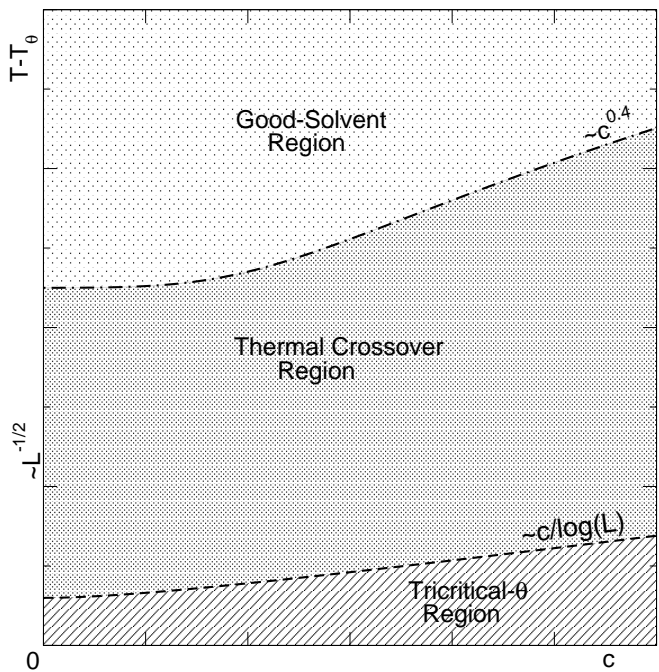,angle=0,width=9truecm} \hspace{0.5truecm} \\
\end{tabular}
\end{center}
\caption{Temperature($T$)-concentration($c$) phase diagram for solutions of polymers of finite length $L$.}
\label{fig:1}
\end{figure}

For finite values of $L$, a gradual transition between the good- and 
$\theta$-solvent regimes is observed, i.e., the observables and their 
associated scaling functions interpolate smoothly between the 
two universal behaviors. A sketch of the various regimes in the phase diagram 
of polymers of finite length is shown in Figure \ref{fig:1}.
At low polymer concentration, i.e., for $\Phi\ll 1$, when increasing the 
temperature above the $\theta$ point, the coil size is observed to 
behave as $L^\nu$ 
with an apparent exponent that increases from $1/2$ to $3/5$. 
Analogously, the 
second virial combination $A_2$ varies smoothly in the range 
$0 \leq A_2 \leq 5.5$, with the value 5.5 (plus finite $L$ corrections) 
obtained in the fully developed good-solvent regime. \cite{CMP-06}
At larger polymer concentrations and for any $T \ge T_{\Theta}$, a 
region where 2-body and 3-body terms are of the same order of magnitude 
is present and is generically referred as {\em tricritical region}. 
For small concentrations this region extends in a small temperature interval
\cite{DaoudJannink,deGennes-79}
whose width is of the order of $1/\sqrt{L}$ (more precisely, 
it  scales as $L^{-1/2} (\ln L)^{-7/11}$). As the concentration $c$ increases,
this region widens as $c/\ln L$.
In the tricritical region scaling 
corrections decay very slowly, as 
inverse powers of $\log L$.\cite{Duplantier,dCJ-book,Schaefer-99,HS-99}
Outside the tricritical region the physics is again dominated by two-body 
effects and by the unbalance between the short-range repulsion and the 
long-range attraction, as it was shown by computer simulations of 
physical polymer models.\cite{PH-05} Here we refer to the 
\emph{thermal crossover region}, as the region in the phase diagram between 
the tricritical and good-solvent regimes. At finite concentration, scaling
 arguments\cite{CMP-08} indicate that the size of the thermal crossover region 
grows with $c$ as $c^{1/(6\nu-2)}\sim c^{0.40}$. 
At sufficiently large temperatures, the good-solvent regime is 
reached at any concentration. 

In the thermal crossover region there are experimental and computer-simulation 
evidences that, for sufficiently large values of $L$,
 global properties of the solution follow, up to small corrections in 
$T, \Phi$, and $L$, general relations of the form 
\begin{equation}
\mathcal{O}(T, c, L) = \alpha_1 \mathcal{O}_G(L,c) 
f_{\mathcal{O}}(z, \Phi),
\label{scaling-gen}
\end{equation}
where $\mathcal{O}_G(L,c)$ is the expression of $\mathcal{O}$ for the 
Gaussian chain, the function $f_{\mathcal{O}}$ is called {\em crossover} 
function, $z = \alpha_2 (T-T_{\theta})L^{1/2}$, and
$\alpha_1$ and $\alpha_2$ are nonuniversal constants that 
embody all chemical details. Theory\cite{Duplantier,dCJ-book,Schaefer-99}
supports the scaling behavior (\ref{scaling-gen}), albeit with a 
slightly different scaling variable. Indeed, the crossover 
limit should be taken by keeping $\alpha_2 (T-T_{\theta})L^{1/2} 
(\ln L)^{-4/11}$ fixed, which differs by a power of $\ln L$ 
from the scaling variable $z = (T-T_{\theta})L^{1/2}$
which appears in Eq.~(\ref{scaling-gen}). Such a logarithmic dependence
is irrelevant in all practical applications, since the observation of this 
slowly varying term would require data in a very large interval of 
polymer lengths/molecular weights. By varying $z$ one obtains 
the full crossover behavior, from the tricritical region, corresponding to 
small values of $z$, to the good-solvent regime, which is obtained for 
$z\to\infty$.

It is important to note that the nontrivial universal behavior is 
obtained when taking simultaneously the limits $L\to\infty$ and 
$T\to T_{\theta}$ in such a way that the arguments of the crossover 
function $f_{\mathcal{O}}$, $z\equiv \alpha_2(T-T_{\theta})L^{1/2}$ and $\Phi$, 
remain constant. If the limits are taken differently, one would obtain 
a different result. For instance, if one takes the limit $L\to \infty$ 
at fixed $T>T_\theta$, one would obtain good-solvent behavior in all
cases, while, if one decreases $T$ towards $T_\theta$ at fixed large $L$, 
only the tricritical behavior would be observed. A similar variety of 
universal scaling behaviors, which depend on how the limit $L\to \infty$ 
is taken, is found in the stretched-chain problem.\cite{Stretched}

The variable $z$ that parametrizes the temperature crossover depends on the 
nonuniversal parameter $\alpha_2$, hence a proper definition requires
specifying a physically meaningful normalization condition. 
To avoid this problem, one can proceed as suggested in
Refs.~\onlinecite{Nickel-91,PH-05},
i.e., one can parametrize the crossover in terms of a 
physical variable rather than in terms of $z$. For
instance, one can use the dimensionless second-virial combination $A_2$.
Then, Eq.~(\ref{scaling-gen})
can be written as
\begin{equation}
{\cal O}(T,L,c) = \alpha_1 {\cal O}_G(L,c) g_{\cal O} (A_2,\Phi),
\label{scaling-gen-2}
\end{equation}
where $g_{\cal O} (A_2,\Phi)$ is universal.
The {\em quality} of the solution is now characterized by $A_2$ that 
varies between zero at the $\theta$ point and $A_{2,GS} \simeq 5.5$, 
the good-solvent value. \cite{CMP-06} 
In the thermal crossover region the relation between $z$ and $A_2$ has been 
obtained combining analytical results  and Monte Carlo simulations, 
\cite{CMP-08} obtaining
\begin{eqnarray}
&& A_2(z) = 
4 \pi^{3/2} z (1 + 19.1187 z 
\label{A2-to-z}
\\
&& \qquad\quad
    + 126.783 z^2 + 331.99 z^3 + 268.96 z^4)^{-1/4},
\nonumber 
\end{eqnarray}
where $z$ has been normalized so that $A_2(z) \approx 4 \pi^{3/2} z$ for 
small $z$, i.e.,~$\Psi(z) \approx z$, where $\Psi = 2 (4 \pi)^{-3/2} A_2$ is 
the so-called interpenetration ratio, which is commonly used in much 
of the experimental literature.

As discussed by Sokal,\cite{Sokal-94} the universal scaling functions that 
parametrize the thermal crossover coincide with the crossover functions 
that are defined in the two-parameters model
(TPM),\cite{Yamakawa-71,dCJ-book,Schaefer-99} which
considers only repulsive two-body monomer-monomer interactions. 
In this framework the scaling variable $z$ is identified with the usual 
Zimm-Stockmayer-Fixmann variable.\cite{ZSF-53}  Therefore, to determine 
the leading crossover behavior, it is convenient to work in the TPM framework, 
which provides directly the crossover functions. This is the approach
we have taken in this paper, using the lattice Domb-Joyce model, \cite{DJ-72}
which is the lattice version of the continuum TPM, to obtain the 
crossover functions, both in the dilute and in the semidilute regime. 

Although the physics of homogenous polymer solutions is 
well understood,
in more complex situations---for instance, inhomogeneous cases or 
when the polymers are only a single component of a more complex system---it 
might be important to reduce the number of degrees of freedom 
representing the polymer subsystem or, equivalently, to limit our 
interest to the physics occurring at a length scale comparable to the 
polymer size. In these cases adopting a coarse-grained (CG)
representation of the solution might be essential for the 
feasibility of the numerical investigation. 
Moreover, since the target properties on which the coarse-grained model (CGM) 
is built  
can be determined in the scaling limit $L\to \infty$, CGMs allow one
to study thermodynamics and structural properties  in the universal,
scaling limit without requiring additional extrapolations.

Several routes to CGMs can be adopted.
In the structure-based route 
\cite{MullerPlathe-02,Reith:2003p2128,PK-09,KVMP-12} 
the CGM is set up in such a 
way to reproduce the marginal probability distribution of a set of chosen 
structural collective variables. This procedure, however, does not 
predict the correct thermodynamics of the underlying solution. 
Conversely, a CGM can be defined to match the thermodynamic behavior, but 
then structural properties are not reproduced 
correctly.\cite{WJK-09} A third method is the force-matching 
approach (often called multiscale coarse-graining method 
\cite{force-matching1}), in which the state-dependent pair-wise 
potential is determined by requiring the CG system to match the 
atomistic force on the CG atoms as accurately as possible. Also this 
method has a structural interpretation:\cite{YBGCG} the matching 
condition is equivalent to require the CG force to satisfy the appropriate 
Yvon-Born-Green equation that relates the pair and the three-body 
correlation function. However, no guarantee of reproducing the 
pair structure and the thermodynamics is provided. The inconsistency 
between structure and thermodynamics stems from the fact that CGMs
neglect multi-body interactions, which would be required 
to obtain an exact mapping of the microscopic model onto the CGM.

Recently, we have introduced a procedure to set up a hierarchy of CGMs for 
linear chains in good-solvent conditions which simultaneously reproduce
quite accurately
structure and thermodynamics of polymer solutions deep into the semidilute 
regime.\cite{DPP-12-Soft,DPP-12-JCP} 
The minimal model consists in representing each linear chain by a short 
polyatomic molecule with four CG sites (tetramer). The tetramer 
potentials are set up at zero density---this allows us to avoid 
the inconsistencies\cite{SST-02,Louis-02,DPP-13-JCP} that occur when
using state-dependent potentials---by matching the single-chain 
intramolecular structure and the center-of-mass pair correlation function 
between two identical chains. The minimal representation (tetramer) has 
been shown to provide accurate results for the underlying solutions 
up to $\Phi\simeq2$.\cite{DPP-12-Soft} Larger values of $\Phi$ can be 
accessed by increasing the resolution of the CGM, i.e. by increasing the 
number of effective monomers (blobs) per chain. The effective 
potentials for these higher-resolution models are obtained by 
using a simple transferability approach. We assume that the potentials
are independent of the number of blobs, as long as all lengths are 
expressed in terms of the blob radius of gyration. This transferability 
approximation was shown to be quite accurate\cite{DPP-12-JCP} and allowed
us to obtain accurate thermodynamic and (large-scale) structural 
results for $\Phi\gg 1$.

In this paper we present the extension of this CG strategy to solutions in the 
thermal crossover region, defining CGMs with $z$-dependent potentials
that reproduce the polymer crossover functions. 
We have considered the TPM as the underlying 
microscopic model, using the lattice Domb-Joyce (DJ) model\cite{DJ-72}
as reference system.
As in the good-solvent case, we first determine
the effective interactions for the tetramer 
case for several values of $z$ in the crossover region,
by matching structural properties at zero density.
These potentials are then used to define CGMs of higher resolutions, i.e.,
higher number of CG sites. The transferability approach is more complex than
in the good-solvent case, since one must change at the same time 
the reference length scale and the scaling variable $z$.
We perform extensive simulations of the CGMs at finite density, both in the 
dilute and in the semidilute region. The results are then compared with 
those obtained in full-monomer (FM) simulations and with the field-theory
predictions of Ref.~\onlinecite{Schaefer-99}. We find that the tetramer CGM  
reproduces very accurately the crossover functions up to $\Phi\approx 30$
(and quite reasonably at all densities)
for small values of $z$. Close to the good-solvent behavior, instead
accurate results are obtained up to $\Phi\simeq 2$. Use of the 
decamer model with $n=10$ blobs allows one to widen significantly
the density region in which the CGM is predictive.

The present work represents an important extension of our previous 
results, allowing us to consider polymer systems in the thermal
crossover regime. In particular, it opens the way to the study of 
more complex systems 
like polymer-colloid solutions away from the good-solvent regime and 
diblock copolymer solutions,\cite{BlockCop,BlockCop1} where each 
block has different affinity with the solvent.

The paper is organized as follows. 
In section \ref{sec2} we define the DJ model 
we have adopted and explain how the crossover functions are computed. 
In section \ref{sec3} we present our CG strategy, 
first illustrating the CG representation (CGR) of the 
FM  chain (section \ref{sec3.1}) and later presenting our 
CGM (section \ref{sec3.2}). Section \ref{sec4} is devoted to the comparison 
between the results of the CGR of the FM system and of the CGM, 
first at zero density (section \ref{sec4.1}) and then at finite density 
in the semidilute regime  (section \ref{sec4.3}). 
Simulation results are also compared with
field-theoretical expressions\cite{Schaefer-99} and large-density predictions 
obtained by using 
the random-phase approximation (RPA), which becomes exact for $\Phi\to \infty$.
In section \ref{sec5},
the transferability of the tetramer potentials to finer resolution models is 
illustrated and validated against FM predictions.
Finally, section \ref{sec6} reports our conclusions. 
Two appendices are also present. In Appendix \ref{App.A} we report the 
calculation of the blob 
radius of gyration in the crossover regime, while in Appendix \ref{App.B} we 
summarize the field-theory predictions of Ref.~\onlinecite{Schaefer-99}. 
Some numerical details and explicit expressions for the CGM potentials 
can be found in the supplementary material.\cite{suppl}

\section{Crossover functions from Monte Carlo Simulations of the Domb-Joyce Model} \label{sec2}

In order to compute the TPM crossover functions,
we consider the three-dimensional lattice Domb-Joyce (DJ)
model.\cite{DJ-72}
In this model the polymer solution is mapped onto $N$ chains of $L$ monomers each 
on a cubic lattice of linear size $M$ with periodic boundary
conditions. Each polymer chain is modelled by a random walk
$\{{\mathbf r}_1^{(i)},\ldots,{\mathbf r}_L^{(i)}\}$ with
$|{\mathbf r}_\alpha^{(i)}-{\mathbf r}_{\alpha+1}^{(i)}|=1$ (we take the
lattice 
spacing as unit of length) and
$1\le i \le N$. The Hamiltonian is given by
\begin{eqnarray}
H &=& \sum_{i=1}^N \sum_{1\le \alpha < \beta \le L}
  \delta({\mathbf r}_\alpha^{(i)},{\mathbf r}_\beta^{(i)}) 
\nonumber \\
&+ &
  \sum_{1\le i < j \le N} \sum_{\alpha=1}^L \sum_{\beta=1}^L
   \delta({\mathbf r}_\alpha^{(i)},{\mathbf r}_\beta^{(j)}),
\end{eqnarray}
where $\delta({\mathbf r},{\mathbf s})$ is the Kronecker delta.
Each configuration is weighted by $e^{-w H}$, where $w > 0$ is a free
parameter that plays the role of inverse temperature.
This model is similar
to the standard lattice self-avoiding walk (SAW) model,\cite{SAW,SAW1} 
which is obtained in the limit  $w \to +\infty$. For finite positive $w$
intersections are possible although energetically penalized.
For any positive $w$, this model has the same scaling limit as the
SAW model\cite{DJ-72} and thus allows us to compute the
universal scaling functions that are relevant for polymer solutions
under good-solvent conditions.

The TPM results in the thermal crossover region can also 
be derived from simulations of the DJ model. 
They are obtained\cite{BD-79} by taking the limit
$w\to 0$, $L\to \infty$ at fixed $x = w L^{1/2}$.
The variable $x$ interpolates between the ideal-chain limit ($x=0$) and
the good-solvent limit ($x=\infty$). Indeed, for $w = 0$ the DJ model is
simply the random-walk model, while for any $w\not=0$ and $L\to \infty$
one always obtains the good-solvent scaling behavior. The variable
$x$ is proportional to the variable $z$ that is used
in the TPM context. If we normalize $z$ so that 
$A_2(z) \approx 4 \pi^{3/2} z$ for small $z$ as in Eq.~(\ref{A2-to-z}), we have
\cite{BD-79,BN-97}
\begin{equation}
 z \equiv \left({3\over 2\pi}\right)^{3/2} w L^{1/2}.
 \label{zdef-DJ}
\end{equation}
As discussed in Ref.~\onlinecite{CMP-08}, the TPM results can be obtained
from Monte Carlo simulations of the DJ model
by properly extrapolating the numerical results to $L\to\infty$. 
For each $z$
we consider several chain lengths $L_i$. For each of them 
we determine the interaction parameter $w_i$ by using Eq.~(\ref{zdef-DJ}),
that is we set
$w_i = (2\pi/3)^{3/2} z L_i^{-1/2}$. Simulations of chains of $L_i$ 
monomers are then performed setting
$w=w_i$. Simulation results are then extrapolated to $L\to \infty$,
taking into account that corrections are of order
$1/\sqrt{L}$.\cite{BD-79,BN-97} If $R(L,z)$ is 
a dimensionless ratio of two global quantities at zero density
(for instance, the 
second-virial combination $A_2$), the TPM result $R^{*}(z)$ is obtained 
by performing an extrapolation of the form
\begin{equation}
R(L,z) = R^*(z) + {a(z)\over \sqrt{L}} + O(L^{-1}\ln L).
\label{extr-R}
\end{equation}
Of course, for $z\to \infty$, $R^*(z)$ converges to its universal good-solvent
value, which can be obtained by taking the limit $L\to \infty$ at fixed 
(arbitrary) $w$.

In this work we often consider 
adimensional distribution functions $g(\rho;L,z)$ that also depend
on the adimensional ratio $\rho = r/\hat{R}_g$. For these properties,
the TPM result
$g^*(\rho;z)$ is obtained by performing an extrapolation of the form
\begin{equation}
g(\rho;L,z) = g^*(\rho;z) + {a(\rho;z)\over \sqrt{L}} + O(L^{-1}\ln L).
\label{extr-V}
\end{equation}
At finite density we should take the limit $L\to \infty$, keeping the polymer 
volume fraction $\Phi$ fixed. In practice we keep 
the dimensionless combination 
$\Phi = 4 \pi c [\hat{R}_g(L_i,w_i)]^3/3$ fixed,
where $\hat{R}_g(L,w)$ is the zero-density radius of gyration (here and in 
the following we indicate any zero-density quantity with a hat).

In order to determine the crossover behavior, we have performed simulations
at the five values $z^{(i)}$, $i=1,\ldots 5$, considered in 
Ref.~\onlinecite{CMP-08}, see Table~\ref{tab-z}. They belong to the
crossover region between ideal and good-solvent behavior
and are such that $A_2(z^{(n)}) \approx n$ (remember that
$A_2(z)$ varies between 0 and 5.50). 
Some additional simulations have also been performed 
at four values of $z$ such that $A_2(z) = 1.5$, $\ldots$, 4.5, where 
$A_2(z)$ is given in Eq.~(\ref{A2-to-z}). The explicit values are 
reported in Table~\ref{tab-z}.

\begin{table}
\caption{Values of $z$ considered in this paper. 
The notation is such that $A_2(z^{(p)}) \approx p$.
The estimates of $A_2(z)$ for $z^{(1)}$, $z^{(2)}$, $z^{(3)}$, $z^{(4)}$,
and $z^{(5)}$,
are the direct MC estimates of Ref.~\protect\onlinecite{CMP-08}; in the other
cases we use the interpolation formula (\protect\ref{A2-to-z}).
In the last column we report $r = A_2(z)/A_{2,GS} = \Psi(z)/\Psi_{GS}$ 
($\Psi$ is the interpenetration ratio
often used in experimental work),
where $A_{2,GS}$ and $\Psi_{GS}$ are the good-solvent values.
}
\label{tab-z}
\begin{tabular}{lclcc}
\hline\hline
\multicolumn{3}{c}{$z$}  & $A_2(z)$  &   $r$ \\
\hline
$z^{(1)}$ &=& 0.056215   & 0.9926(10) & 0.18 \\
$z^{(1.5)}$ &=&0.097563  & 1.5        & 0.27 \\ 
$z^{(2)}$ &=& 0.148726   & 1.9782(18) & 0.36 \\
$z^{(2.5)}$ &=&0.225292  & 2.5        & 0.45 \\ 
$z^{(3)}$ &=& 0.321650   & 2.9621(27) & 0.54 \\ 
$z^{(3.5)}$ &=&0.493088  & 3.5        & 0.64 \\ 
$z^{(4)}$ &=& 0.728877   & 3.9433(34) & 0.72 \\
$z^{(4.5)}$&=&1.32527    & 4.5        & 0.82 \\
$z^{(5)}$ &=& 2.50828    & 4.9147(36) & 0.89 \\
\hline\hline
\end{tabular}
\end{table}

In principle, the crossover functions can also be computed in other 
models that interpolate between good-solvent and $\theta$ behavior.
A typical example, which has been widely discussed in the 
literature, see, e.g., Refs.~\onlinecite{Grassberger1,KHL-03,ALH-04}, is 
the SAW model with nearest-neighbor interactions (often called interacting 
SAW model). Another model, quite interesting from a computational 
point of view, is the extension of the DJ model discussed in 
Ref.~\onlinecite{HS-99}, in which an additional energy term associated with 
triple intersections is considered.
Unlike the DJ model, in these models there is a nonvanishing three-body 
effective coupling, which implies, for instance, that 
the third virial combination $A_3$ is positive at the 
Boyle temperature where $A_2 = 0$. If we take these marginally irrelevant 
three-body terms into account, the scaling behavior (\ref{scaling-gen}) 
becomes \cite{dCJ-book,Schaefer-99,HS-99} 
\begin{equation}
{\cal O}(T,L,c) = \alpha_1 O_G(L,c) h_{\cal O}(z,u_3,\Phi),
\label{scaling-gen-u3}
\end{equation}
where $u_3$, which parametrizes the effective three-body interaction, 
vanishes as $1/\ln L$ for $L\to \infty$. 
As long as the thermal crossover region is considered, i.e., we are 
in the temperature/degree-of-polymerization region such that $z\gg u_3$
and $L$ is large,
we can neglect $u_3$ in Eq.~(\ref{scaling-gen-u3}) and reobtain
Eq.~(\ref{scaling-gen}) with $f_{\cal O}(z,\Phi) = h_{\cal O}(z,0,\Phi)$.
However, for polymers of finite length, tricritical corrections proportional
to $u_3$ give rise to slowly varying scaling corrections, which make a precise
determination of the scaling crossover functions quite difficult.
For this reason it is convenient to consider models, 
like the DJ one, with only two-body repulsion, thereby 
avoiding unwanted tricritical corrections.
Of course, if one wishes to discuss also tricritical effects, the model of 
Ref.~\onlinecite{HS-99} or interacting SAWs should be considered. In this
respect, we note that the extended DJ model\cite{HS-99} is 
computationally much more convenient than the more common interacting SAW 
model. Indeed, since interactions are soft, one expects the Monte Carlo 
dynamics to be significantly faster than for interacting SAWs.

\section{The blob model} \label{sec3}

\subsection{The coarse-grained 
representation of the polymer model} \label{sec3.1}

In the multiblob approach one starts from a {\em coarse-grained representation}
(CGR) of the underlying full-monomer (FM) model, which is obtained by
mapping a chain of $L$ monomers onto a chain of $n$ blobs, each of them
located at the center of mass of a subchain of $m=L/n$ monomers. 
If the monomer positions are given by
$\{ {\bf r}_1,\ldots, {\bf r}_L\}$, one first defines the
blob positions ${\bf s}_1,\ldots, {\bf s}_n$ as the
centers of mass of the subchains of length $m$, i.e.
\begin{equation}
 {\bf s}_i = {1\over m} \sum_{\alpha=m(i-1)+1}^{mi}  {\bf r}_\alpha.
\end{equation}
For the new CG chain $\{{\bf s}_1,\ldots, {\bf s}_n\}$
one defines several standard quantities. First, one defines its
square radius of gyration
\begin{equation}
{R}_{g,b}^2(n) = {1\over 2 n^2} \sum_{i,j=1}^n ({\bf s}_i - {\bf s}_j)^2 .
\label{Rgb-def}
\end{equation}
Such a quantity is always smaller than ${R}_g^2$, because of the exact identity
\begin{equation}
{R}_g^2 = {R}_{g,b}^2(n) + r^2_g(n),
\label{Rg-Rgb}
\end{equation}
where ${r}_{g}(n)$ is the average radius of gyration of the blobs.
The ratios $R_{g,b}^2(n)/R_g^2$ and $r_{g}^2(n)/R_g^2$ of their averages
over the polymer configurations (we use the same symbol both for the 
radius of a single chain and for its average; the correct interpretation
should be clear from the context) show a universal crossover behavior, 
i.e. independent
of the nature of the underlying polymer model as long as $L$ and $m$ are
large enough. 
This crossover can be equivalently parametrized in terms of $z$ or of $A_2$. 
Explicit zero-density results  are reported in App.~\ref{App.A}.

To define the CGM we proceed as in our 
previous work,\cite{DPP-12-Soft,DPP-12-JCP} 
defining at first a tetramer model with $n=4$ blobs.
Higher-resolution models with $n > 4$ will be discussed in Sec.~\ref{sec5}. 
To determine the four-blob CGM, we compute
several intramolecular CGR structural distributions for an isolated chain
(zero-density limit).
First, we determine
the bond-length distributions of the CGR FM model:
\begin{equation}
P_{ij}(r) = \langle \delta(|{\bf s}_i - {\bf s}_j| - r) \rangle.
\end{equation}
In the crossover limit at fixed $z$, the adimensional combination 
$\hat{R}_g P_{ij}(r)$ converges to a universal crossover function 
$f_{ij}^*(\rho;z)$, $\rho = r/\hat{R}_g$, which can be computed in the 
DJ model as described in Sec.~\ref{sec2}. 
Second, we will need
the distributions of the two equivalent bending angles $\beta_i$, 
of the torsion angle $\theta$, and of the angle $\beta_{13}$, defined 
in the CGR of the polymer model. They are defined as 
follows: 
\begin{eqnarray}
&& \cos\beta_i = -{{\bf b}_i \cdot {\bf b}_{i+1} \over 
              |{\bf b}_i| |{\bf b}_{i+1}|},
\label{bending} \\
&& \cos\beta_{13} = {{\bf b}_1 \cdot {\bf b}_3 \over 
              |{\bf b}_1| |{\bf b}_3|},
\label{bending13} \\
&& \cos\theta = 
   {({\bf b}_1 \times {\bf b}_2) \cdot 
    ({\bf b}_2 \times {\bf b}_3) \over 
   |{\bf b}_1 \times {\bf b}_2|
   |{\bf b}_2 \times {\bf b}_3| },
\label{torsion}
\end{eqnarray}
with ${\bf b}_i = {\bf s}_{i+1} - {\bf s}_i$. In the crossover limit
they converge to the TPM distributions $f_b^*(\cos\beta;z)$,
$f_{b,13}^*(\cos\beta_{13};z)$, and $f_t^*(\theta;z)$. 

Finally, to determine the intermolecular CGM interactions we will make use of 
the center-of-mass intermolecular distribution function.
It is defined by
\begin{equation}
g_{CM}(r) = 
\langle e^{-\beta U_{12}} \rangle_{0,\bf r},
\end{equation}
where $\langle\cdot \rangle_{0,\bf r}$ indicates the average over two isolated
polymers, the centers of mass of which are in the origin and in
$\bf r$, respectively, and $U_{12}$ is the intermolecular energy.
In the crossover limit $L\to\infty$ at fixed $z$, $g_{CM}(r)$ converges to
a universal function $g^*_{CM}(\rho;z)$.

\subsection{The coarse-grained blob model} \label{sec3.2}

The CGM consists of polyatomic molecules of $n$ atoms located in 
$\{{\bf t}_1,\ldots,{\bf t}_n\}$. All length scales are expressed in terms 
of the zero-density radius of gyration, hence all potentials and 
distribution functions depend on the adimensional combinations 
${\bm \rho} = {\bf t}/\hat{R}_g$. 
In order to have an exact mapping of the CGR of the polymeric system onto the
$n$-blob CGM, one should consider an $n$-body
intramolecular potential, which, for $n>2$,
can be expressed in terms of $3(n-2)$ scalar combinations of
the positions of the blobs because of rotational and translational invariance.
Even for $n$ as small as 4, this requires considering a function of
6 independent variables.
Moreover, since we are computing the potentials in the crossover 
region, we should additionally consider them as a function of the crossover 
variable $z$.
Of course, this is far too complex in practice. Hence, we have 
used two different simplifications.\cite{DPP-12-Soft,DPP-12-JCP} First, 
we use a limited set of interactions.
The intramolecular interactions have been modeled by introducing
seven different potentials, each of them depending on a single scalar variable.
This choice is arbitrary, but, as we have already verified in the 
good-solvent case,\cite{DPP-12-Soft,DPP-12-JCP} it is
particularly convenient and works quite well.
Second, we have computed the potentials only for five different values of $z$,
$z^{(1)}$, $\ldots$, $z^{(5)}$, 
reported explicitly in Table~\ref{tab-z}. For other values of $z$ we use
a simple interpolation, which, as we shall discuss, works quite precisely.

As in Refs.~\onlinecite{DPP-12-Soft,DPP-12-JCP},
 we consider  a set of bonding pair potentials:
blobs $i$ and $j$ of the tetramer interact with a pair potential
$V_{ij}(\rho;z)$ with $\rho = |{\bf t}_i - {\bf t}_j|/\hat{R}_g$.
Because of symmetry we have
$V_{13}(\rho;z) = V_{24}(\rho;z)$ and $V_{12}(\rho;z) = V_{34}(\rho;z)$, so that
there are only four independent potentials to be determined.
Then, we consider
a bending-angle potential $V_b(\cos \beta;z)$, 
a potential $V_{b,13}(\cos\beta_{13};z)$
and a torsion-angle potential $V_t(\theta;z)$, where 
$\beta$, $\beta_{13}$, and $\theta$ are defined as in Sec.~\ref{sec3.1}.

To determine the CGM intramolecular potentials, we require the CGM to 
reproduce the adimensional TPM bond distributions $f^*_{ij}(r;z)$ and the 
TPM angle distributions defined in Sec.~\ref{sec3.1}.  To obtain these 
distributions, we perform simulations of the DJ model with chains of 
lengths $L=1000,2500$, and 5000 monomers, using the pivot algorithm.
\cite{Lal,MacDonald,Madras-Sokal,Sokal-95b,Kennedy-02,Clisby-10b}
Then, we extrapolate the relevant 
quantities by using Eqs.~(\ref{extr-R}) and (\ref{extr-V}). 
Once the TPM distributions are known, the potentials are obtained 
by applying the Iterative Boltzmann Inversion (IBI) scheme.
\cite{Schommers:1983p2118,MullerPlathe-02,Reith:2003p2128}

\begin{figure}[t]
\begin{center}
\begin{tabular}{c}
\epsfig{file=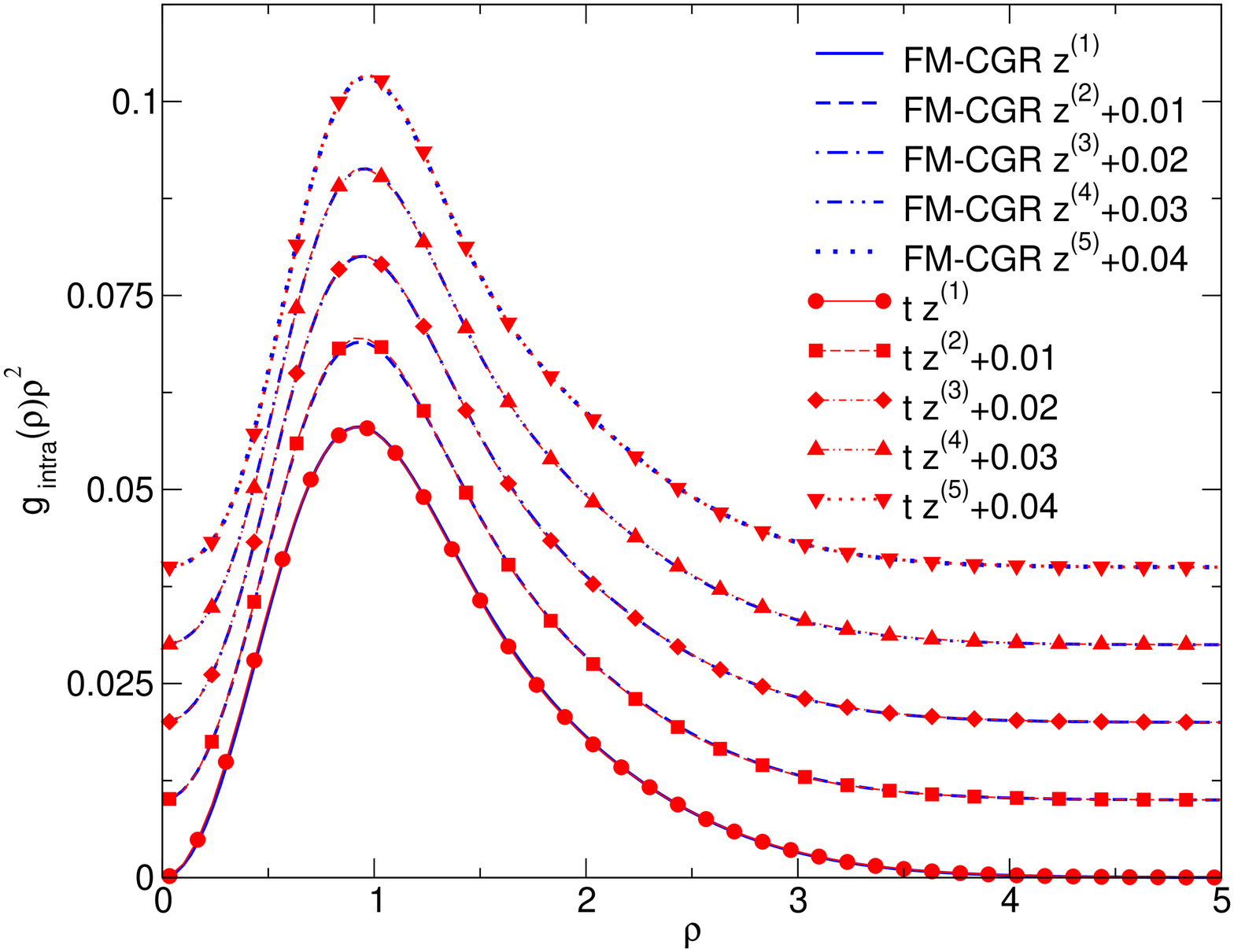,angle=0,width=9truecm} \hspace{0.5truecm} \\
\end{tabular}
\end{center}
\caption{Intramolecular distribution function $\rho^2 g_{\rm intra}(\rho;z)$. 
We report tetramer (t) and full-monomer (FM) results at zero density for 
the values of $z$ reported in Table~\ref{tab-z}.
For sake of clarity, results at different values of $z$
are shifted upward according to the legend.
}
\label{fig:gintra}
\end{figure}

\begin{table}[t]
\caption{Estimates of $\hat{R}^2_{g,b}/\hat{R}^2_g$ and of $A_2$ 
obtained by using the tetramer (t) CGM and the full-monomer (FM)
model. The values of $z$ are reported in Table~\protect\ref{tab-z}. }
\label{table:A2-Rgb}
\begin{tabular}{ccccc} 
\hline\hline
& 
\multicolumn{2}{c}{$\hat{R}^2_{g,b}/\hat{R}^2_g$}  & 
\multicolumn{2}{c}{$A_2=B_2/\hat{R}_g^3$} \\
 \hline
 $z$ &  FM  & t  &  FM &  t  \\
\hline
$z^{(1)}$ &  0.7553(7) & 0.7574(3) & 0.9926(10) & 0.9763(1)  \\
$z^{(2)}$ &  0.761(1)  & 0.7587(5) & 1.9782(18) & 1.9459(3) \\
$z^{(3)}$ &  0.7686(15)& 0.7719(5) & 2.9621(27) & 2.9364(5) \\
$z^{(4)}$ &  0.777(2)  & 0.7844(4) & 3.9433(34) & 3.9999(7) \\
$z^{(5)}$ &  0.787(2)  & 0.7885(4) & 4.9147(36) & 4.9105(8) \\
\hline\hline
\end{tabular}
\end{table}

The method works quite precisely: The CGM reproduces quite 
well the target distribution functions. For instance, in Fig.~\ref{fig:gintra}
we report the intramolecular distribution function $g_{\rm intra}(\rho)$ 
computed by using 
the CGR of the polymer model and the tetramer model for the five values of 
$z$ we consider. The agreement is excellent. Excellent agreement
is also observed for the angle distributions (see supplementary 
material\cite{suppl}).
As a second check of the accuracy of the inversion procedure,
we compute the ratio $\hat{R}_{gb}/\hat{R}_g$
both in the polymer model and in the CGM.
Again, the results reported in Table~\ref{table:A2-Rgb} show an 
excellent agreement.

\begin{figure*}[t!]
\centering
\begin{tabular}{c}
\includegraphics[width=13truecm,keepaspectratio,angle=0]{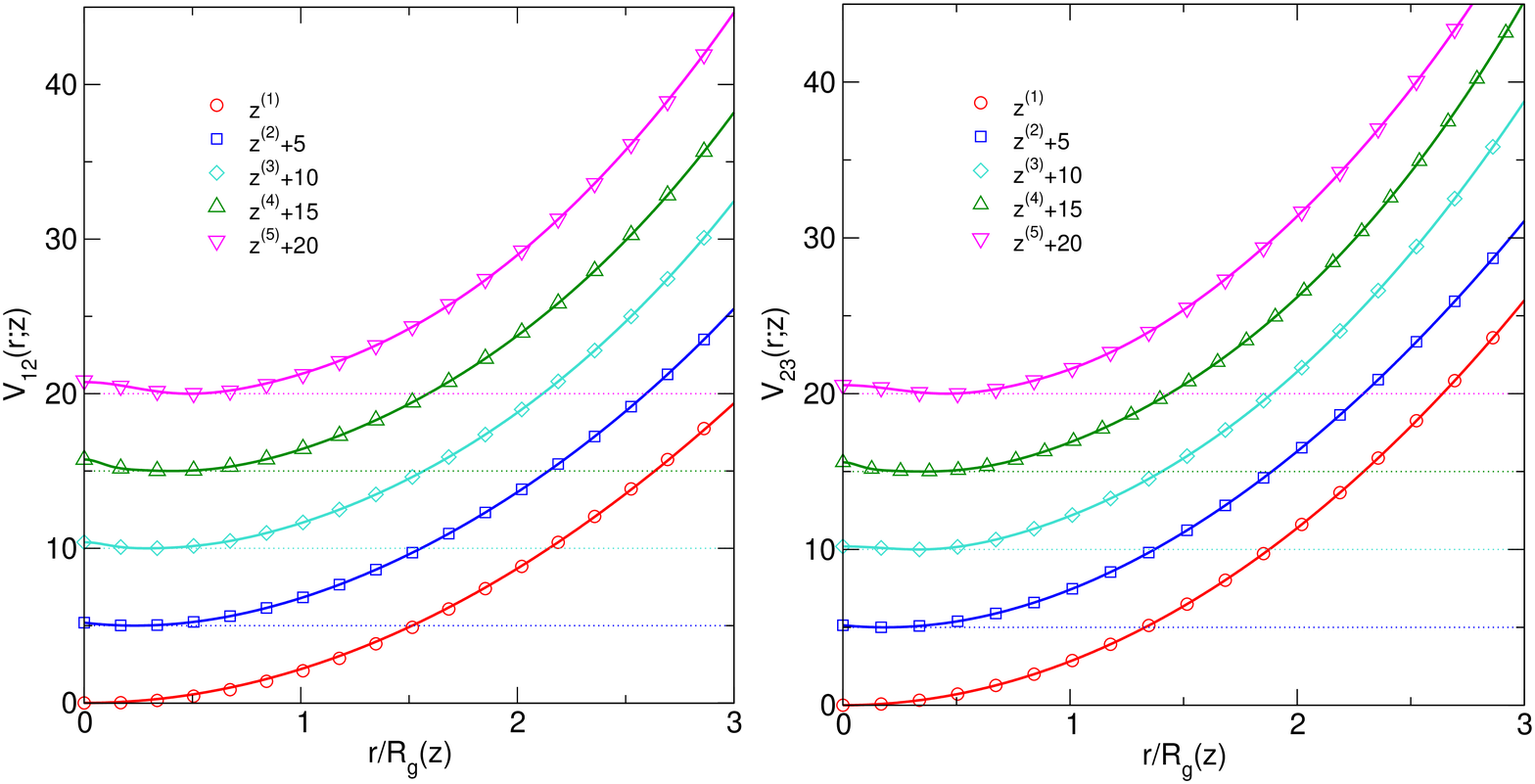}
\\
\includegraphics[width=13truecm,keepaspectratio,angle=0]{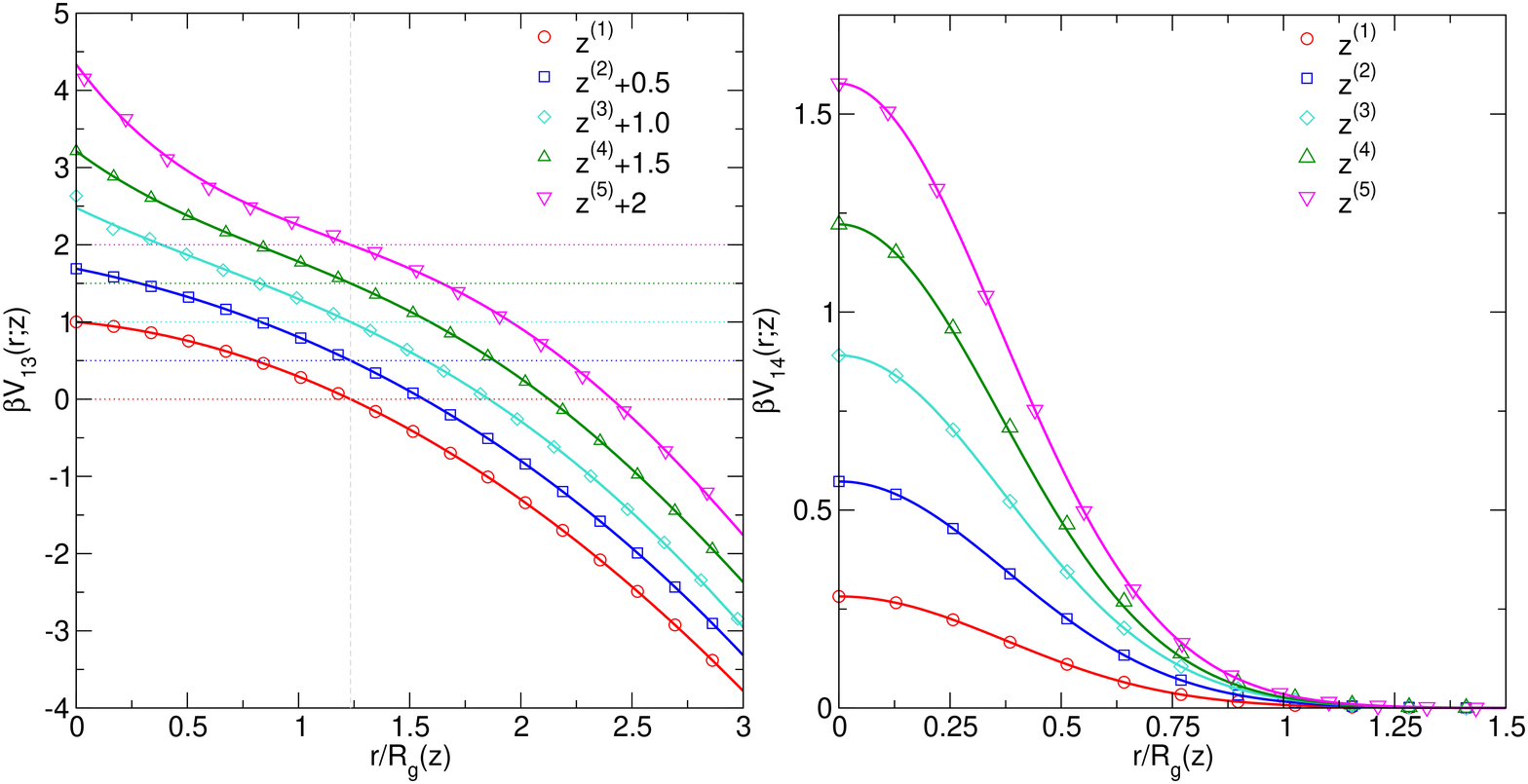}
\\
\end{tabular}
\caption{Intramolecular tetramer bond potentials 
for five values of $z$ reported in Table~\ref{tab-z}: 
points correspond to the numerical estimates, while the solid lines
are the interpolations reported in the supplementary material.\cite{suppl}
For the sake of clarity, results
at different  values of $z$ are shifted upward according to the legend.  }
\label{fig:Intramoleculart1}
\end{figure*}

\begin{figure*}[t!]
\centering
\begin{tabular}{c}
\includegraphics[width=13truecm,keepaspectratio,angle=0]{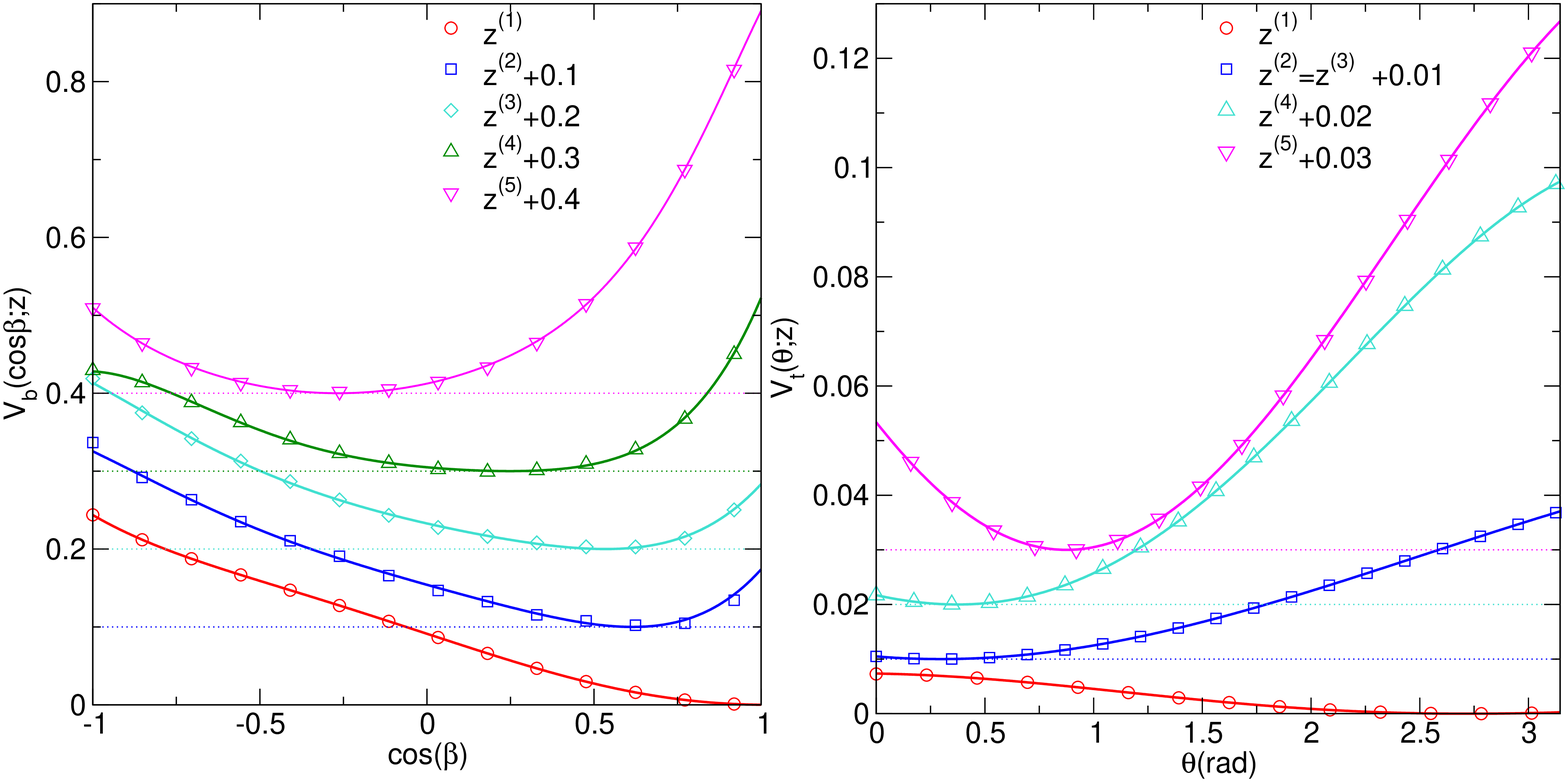}
\\
\includegraphics[width=13truecm,keepaspectratio,angle=0]{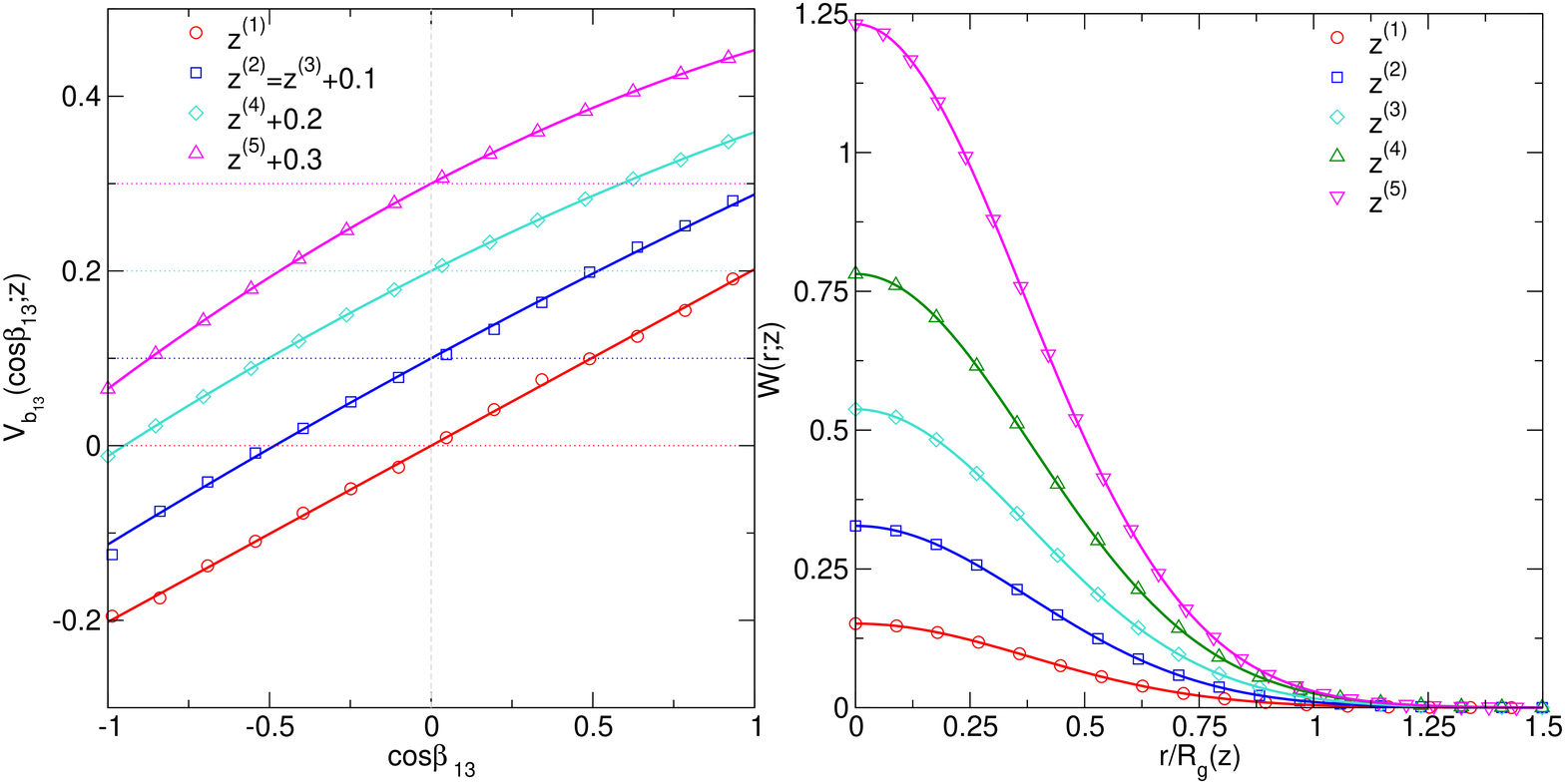}
\\
\end{tabular}
\caption{Tetramer angular and intermolecular potentials 
for five values of $z$ reported in Table~\ref{tab-z}: 
points correspond to the numerical estimates, while the solid lines
are the interpolations reported in the supplementary material.\cite{suppl}
For the sake of clarity, results
at different  values of $z$ are shifted upward according to the legend. 
}
\label{fig:Intramoleculart2}
\end{figure*}

The potentials are shown in Figs.~\ref{fig:Intramoleculart1} and 
\ref{fig:Intramoleculart2}. Explicit parametrizations are reported in the
supplementary material.\cite{suppl}
The bonding pair potentials $V_{12}(\rho;z)$ and $V_{23}(\rho;z)$ change 
significantly with $z$. First, the position $\rho_{\rm min}$ of the minimum of 
the potentials decreases from the good-solvent value\cite{DPP-12-Soft} 
$\rho_{\rm min} = 0.5$ to 0.3 for $z = z^{(3)}$ and is approximately
zero for $z^{(1)}$. This is consistent with what one finds\cite{LOS-91}
for the random-walk case ($z=0$), for which the bonding pair potential 
is a parabola, hence 
$\rho_{\min} = 0$. Second, the potential becomes softer
at the origin as $z$ decreases. For instance, 
$\Delta_{12}(z)=\beta[V_{12}(0;z)-V_{12}(\rho_{\rm min};z)]$ decreases from 
the good-solvent value 0.8 to 0 as $z$ decreases.
More precisely, $\Delta_{12}(z)\approx0, 0.4, 0.7$ for 
$z=z^{(1)},z^{(3)},z^{(5)}$, respectively.
As before, this is due to the fact that excluded-volume effects become 
less relevant as $z$ decreases.
Potential $V_{13}(\rho;z)$ shows qualitatively the same behavior as 
a function of $z$. Potential $V_{14}(\rho;z)$ has an approximate 
Gaussian shape for all values of $z$. As expected $V_{14}(0;z)$ 
decreases significantly with decreasing $z$. Since excluded-volume 
effects decrease in this limit, overlaps are less penalized.

In Fig.~\ref{fig:Intramoleculart2} we report the angular potentials.
The bending potential changes significantly with $z$. For large $z$
folded configurations with $\cos\beta \approx 1$ are penalized, 
while for small $z$, they are slightly favored.
The torsion potential becomes  very small for small values of $z$:
the tetramer becomes more flexible as the excluded-volume interactions become
less effective. For instance $\beta [V_{t}(\pi;z^{(1)})-V_t(0;z^{(1)}]\approx
0.01$.
The four-body bending potential $V_{b,{13}}(\cos\beta_{13};z)$ changes
only in the region in which $\cos\beta_{13}>0$. In particular, by
increasing $z$, $\Delta_{b,{13}}=\beta
[V_{b,{13}}(1;z)-V_{b,{13}}(0;z)]$ decreases, signalling that 
more elongated conformations are preferred as 
excluded-volume effects increase.

\begin{figure}[tbp]
\begin{center}
\begin{tabular}{c}
\epsfig{file=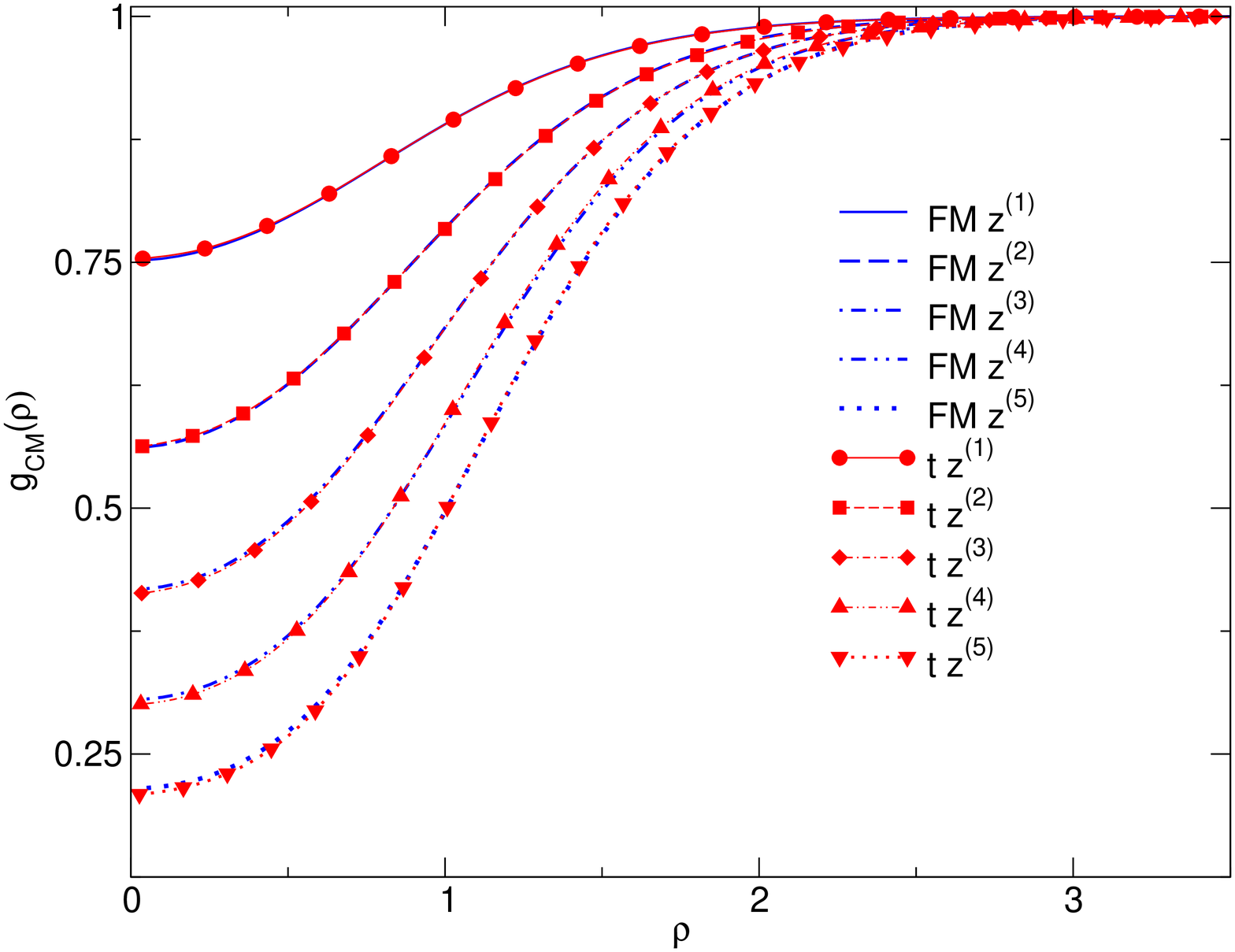,angle=0,width=9truecm} \hspace{0.5truecm} \\
\end{tabular}
\end{center}
\caption{Radial distribution functions for 
a pair of isolated chains  as a function of  the center-of-mass 
separation $\rho = r/\hat{R}_g$. We report tetramer (t) and full-monomer (FM) 
results at zero density for five values of $z$ reported in 
Table~\protect\ref{tab-z}. }
\label{fig:grcm}
\end{figure}

As for the intermolecular potentials, we have made again some drastic 
simplifications. First, in the spirit of the multiblob approach,
we neglect interactions among three or more tetramers. 
The potential between two tetramers is still a function of the $6(n-1)$
relative positions of the blobs, which is again far too complex.
We have thus simplified the model by considering
a single intermolecular central pair potential $W(\rho;z)$:
all blobs interact with the same potential, irrespective of their positions
along the tetramer. As shown in Refs.~\onlinecite{DPP-12-Soft,DPP-12-JCP} 
this drastic simplification works quite well. 
Such a potential has been obtained by requiring
the CGM to reproduce the TPM center-of-mass intermolecular distribution 
function $g^*_{CM}(\rho;z)$.
The potential $W(\rho;z)$ has been parametrized as
\begin{equation}
\beta W(\rho;z) =c_1(z) \exp (-c_2(z) \rho^2), 
\label{Wdef}
\end{equation}
in terms of two unknown $z$-dependent parameters $c_1(z)$ and $c_2(z)$.
They have been determined following the approach of 
Ref.~\onlinecite{Akkermans:2001p1716}. Parametrization (\ref{Wdef})
looks adequate. The CGM reproduces quite precisely the 
FM result, see Fig.~\ref{fig:grcm}.
As a check of the accuracy of the procedure, 
we also compare the second-virial combination $A_2$. 
Again, the results reported in Table~\ref{table:A2-Rgb} show 
good agreement (differences are less than 2\%).
The intermolecular potential is shown in Fig.~\ref{fig:Intramoleculart2}. 
As expected, in the crossover region from good-solvent to $\theta$ behavior
one observes a decrease of its strength and a 
slight decrease of its spatial range.

The results reported above give the tetramer potentials for five different
values of $z$, each of which corresponds to a different
value $A_{2}^{(n)}$ of the second-virial combination $A_2$,
see Table~\ref{tab-z}. To define the model for all values of $z$
we use a simple linear interpolation formula. 
Given $z$, we first determine $A_2(z)$ by using Eq.~(\ref{A2-to-z}) and then 
an integer $1 \le n_1 \le 5$ such that 
$A^{(n_1)}_2 \le A_2(z) \le A_2^{(n_1+1)}$ 
($n=6$ corresponds to the good-solvent case, hence $A_2^{(6)} = 5.500$). 
Then, for each potential we set
($\sigma$ may be $\rho$ or an angular variable)
\begin{eqnarray}
   V(\sigma;z) &=& {A_2(z) - A^{(n_1)}_2\over A_2^{(n_1+1)} - A^{(n_1)}_2} 
          V(\sigma;z^{(n_1+1)}) 
\nonumber \\ &- &
      {A_2(z) - A^{(n_1+1)}_2\over A_2^{(n_1+1)} - A^{(n_1)}_2} 
          V(\sigma;z^{(n_1)}).
\label{VA2-interp-testo}
\end{eqnarray}
As a check of the accuracy of this interpolation we have 
chosen four values of $z$ such that $A_2 = 1.5$, 2.5, 3.5, 4.5, 
see Table~\ref{tab-z}.
For these values we have computed the tetramer potentials using 
the interpolation formula (\ref{VA2-interp-testo}). 
Then, we have computed again $A_2$ for each value of $z$ by using the 
tetramer CGM.
Results are consistent---differences are at most 1\%---confirming the 
accuracy of the interpolation we use (see supplementary material\cite{suppl} 
for more details).

\section{Comparing tetramer and full-monomer predictions} \label{sec4}

In this section, we compare the CGM structural predictions with those 
obtained for the CGR of the polymer model. In Sec.~\ref{sec4.1} 
we extend the discussion of Sec.~\ref{sec3.2} at zero density, while 
in Sec.~\ref{sec4.3} we consider the semidilute regime. Here we 
compare the CGM results with the results of FM
simulations at finite density, discussed in Sec.~\ref{sec4.2}.
Beside considering the tetramer model, we shall also discuss the 
simpler single-blob (SB) model, in which each polymer is 
represented by a monoatomic molecule located in the 
polymer center of mass. The SB potentials in the 
$\theta$-to-good-solvent crossover regime were computed in 
Refs.~\onlinecite{KHL-03,ALH-04,PH-05} using the 
self-avoiding walk model. Here we perform the same computation more 
carefully, by using the DJ model. An explicit parametrization 
that satisfies all theoretical constraints and reproduces the good-solvent
results of Ref.~\onlinecite{PH-05} is reported in the supplementary material.
\cite{suppl}

\subsection{Zero-density results} \label{sec4.1}

\begin{figure}[tb]
\begin{center}
\begin{tabular}{c}
\epsfig{file=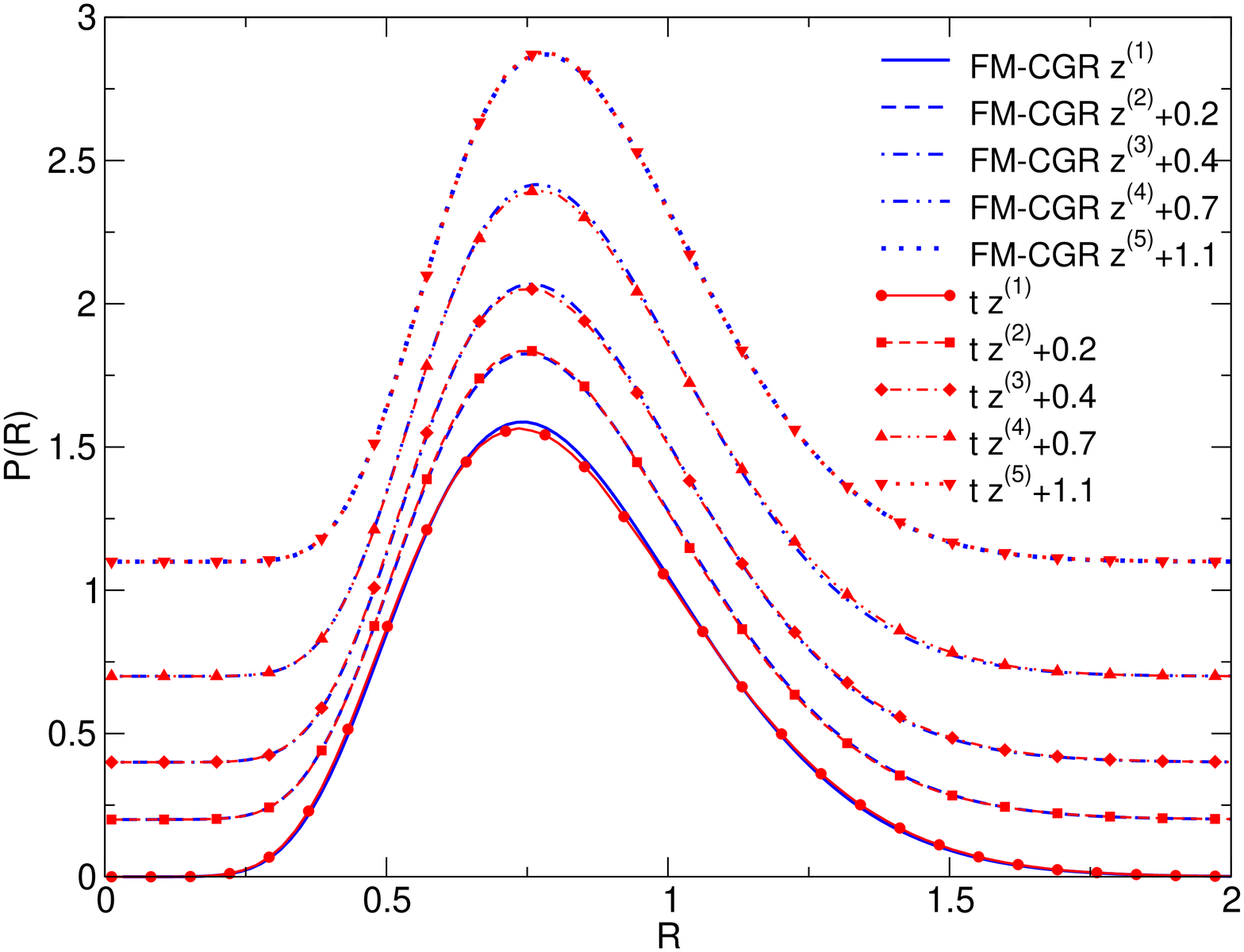,angle=0,width=9truecm} \hspace{0.5truecm} \\
\end{tabular}
\end{center}
\caption{Distribution  of the ratio 
$R=R_{g,b}/\sqrt{\langle\hat{R}^2_g\rangle}$. 
We report tetramer (t) and full-monomer (FM) results at zero density 
for the values of $z$ reported in Table~\ref{tab-z}. 
For the sake of clarity, results at different 
values of $z$ are shifted upward according to the legend.
}
\label{fig:Rgb-dist-zerodens}
\end{figure}

By construction, the tetramer CGM reproduces all CGR bond-length
distributions. However, it is far from obvious that the distributions of other
structural intramolecular quantities are correctly reproduced.
In Fig.~\ref{fig:Rgb-dist-zerodens} we report the distribution of the 
ratio $R = \hat{R}_{g,b}/\langle \hat{R}_g^2\rangle^{1/2}$,
where $\hat{R}_{g,b}$ is the radius associated with a single chain,
while $\langle \hat{R}_g^2\rangle$ is the average squared radius of 
gyration. The 
agreement is excellent for all values of $z$ we consider. 
This is a nontrivial check of the quality of the multiblob representation
since this distribution is not directly related to the bond-length
distributions
nor to those of the intramolecular angles we consider. Clearly, the tetramer
CGM correctly models shape and size of the polymer in the whole 
crossover region.

\begin{table}[t]
\caption{Estimates of $A_3$ for the full-monomer (FM) model,
for the tetramer (t), and single-blob (SB) CGMs. 
We consider values $z^{(n)}$ defined in Table~\protect\ref{tab-z}.
The full-monomer (FM) results for $z^{(1)}$, $z^{(2)}$, $z^{(3)}$, $z^{(4)}$, 
and $z^{(5)}$,
are the direct MC estimates of Ref.~\protect\onlinecite{CMP-08}; in the other 
cases we use the interpolation formula $A_3(z)$ reported in 
Table III of Ref.~\protect\onlinecite{CMP-08}.}
\label{A3-CGBM}
\begin{tabular}{cccc} 
\hline \hline
 $z$ &  FM  & t  &  SB \\
\hline
$z^{(1)}$ &  0.0849(8) & 0.077(1)& 0.059 \\
$z^{(1.5)}$ & 0.276  & 0.261(6)& 0.197 \\
$z^{(2)}$ &  0.6061(30)& 0.553(8)& 0.442 \\
$z^{(2.5)}$ & 1.15   & 1.10(2) & 0.867 \\
$z^{(3)}$ &  1.843(8)  & 1.74(2) & 1.404 \\
$z^{(3.5)}$ & 2.89   & 2.85(3) & 2.264 \\
$z^{(4)}$ &  4.021(13) & 3.95(4) & 3.164 \\
$z^{(4.5)}$ & 5.72   & 5.67(5) & 4.569 \\
$z^{(5)}$ &  7.243(22) & 7.02(6) & 5.827 \\
\hline\hline
\end{tabular}
\end{table}

It is also of interest to compare the predictions for the virial coefficients
defined in Eq.~(\ref{eq:virial}).
Since  we matched the pair distribution function $g_{CM}(r)$, the combination
$A_2 = B_2 \hat{R}_g^{-3}$ should be the same in the CGM and in the polymer 
model, as we have already verified in Sec.~\ref{sec3.2}. 
Here we compare the third-virial combination $A_3 = B_3 \hat{R}_g^{-6}$. 
Differences between the CGM and the polymer results allow us to quantify
the importance of the three-polymer interactions that are not fully
taken into account by the CGM. Beside tetramer and 
FM results, we also report estimates obtained by using 
the SB model.  The results are reported in Table~\ref{A3-CGBM}. The SB results 
underestimate quite significantly the FM results: the relative 
deviations increase from 19\% for $z = z^{(5)}$ to 30\% for 
$z = z^{(1)}$. The tetramer results are significantly better. 
Close to the good-solvent regime ($z\gtrsim z^{(4)}$) deviations are 
approximately 3\%. As $z$ is lowered, differences increase: 
for $z = z^{(1)}$, we find a 9\% difference. It is important to note 
that, although in all cases the observed differences for $A_3$ increase
as the $\theta$ point is approached, this does not imply that 
the difference between the CGM and FM pressure increases 
when lowering $z$. Indeed, perturbation theory gives
$A_n/A_2 \approx z^{n-1}$ for $z \to 0$.
Therefore, the third and higher-order virial contributions to the pressure 
become increasingly less relevant as $z$ decreases. 
As a consequence, as we discuss in Sec.~\ref{sec4.3}, the CGM provides
increasingly more accurate estimates of the pressure as the $\theta$ point 
is approached.

\subsection{Intermezzo: finite-density 
         thermodynamics by full-monomer simulations} \label{sec4.2}

In order to discuss the behavior of the CGM in the semidilute regime, 
we need to obtain FM reference data. For this reason we have 
performed an extensive study of the DJ model at finite density. 
For the five values of $z$, $z^{(1)}$, $\ldots$, $z^{(5)}$, reported in 
Table~\ref{tab-z} we have performed simulations for several values of 
$\Phi$ and several values of $L$ ranging between 100 and 2000.
As in our previous work,\cite{Pelissetto-08}
we use a combination of 
pivot,\cite{Lal,MacDonald,Madras-Sokal,Sokal-95b,Kennedy-02}
cut-and-permute,\cite{CPS-90,Causo-02,Pelissetto-08}
 and reptation moves. Since the penalty parameter
$w$ is quite small, the algorithm is quite efficient and 
for $z^{(1)}$, $\ldots$, $z^{(4)}$ we are able to simulate polymer systems
for densities 
up to $\Phi = 30$. For $z=z^{(5)}$, i.e. close to the good-solvent regime,
the largest density considered corresponds to $\Phi = 20$. In the
finite-density simulation we should also fix the volume $V=M^3$ of the box. 
As in Ref.~\onlinecite{Pelissetto-08} we fix the box size so that the 
number $N$ of polymers in the box is never smaller than 100 (typically 
$150\lesssim N \lesssim 1000$). We measure the relevant CGR intramolecular and
intermolecular distribution functions, needed for a detailed comparison
with the CGM results. Moreover, to compare the thermodynamics we compute 
\begin{equation}
K = {\partial \beta \Pi\over \partial c},
\end{equation}
where $c = N/V$. This quantity is determined as in
Ref.~\onlinecite{Pelissetto-08}.
Using the compressibility rule \cite{HMD-06}
we can relate $K$ to the total structure
factor that can be easily measured in simulations.
If we define
\begin{equation}
S({\mathbf k}) \equiv  {1\over L^2 N}
   \left\langle \left| \sum_{j=1}^N \sum_{\alpha=1}^L 
      \exp (i {\mathbf k}\cdot {\mathbf r}_\alpha^{(j)}) \right|^2
    \right\rangle,
\end{equation}
then we have
\begin{equation}
   {1\over K}  = \lim_{k\to 0} S({\mathbf k}).
\end{equation}
Since we work in a finite box of volume $M^3$, we must quantify the 
finite-volume effects. For fluids they have been extensively discussed,
see, e.g., Refs.~\onlinecite{FSS} and references therein.
In general, at fixed $\Phi$  finite-volume quantities converge to 
their infinite-volume counterpart (if we consider 
distribution functions or the structure factor
we should, of course, take the limit at fixed $r$ or $k$) with corrections
of order $M^{-3}$. We find that these corrections
are smaller than or, at most, of the same order as the statistical
errors. To give an example, let us discuss in more detail the structure factor 
$S({\bf k})$. If we assume 
\begin{equation}
S({\bf k};M) = S({\bf k};\infty) + {\alpha({\bf k})\over M^3} + O(M^{-6}),
\end{equation}
then, we have 
\begin{equation}
S({\bf k};M) - S({\bf k};\infty) =
    {1\over 7} \left[S({\bf k};M/2) - S({\bf k};M)\right] + O(M^{-6}).
\label{Sk_FSS}
\end{equation}
Hence, a rough estimate of the size effects can be obtained 
by computing the same quantity for two boxes of linear size $M$ and $M/2$,
respectively. To give an idea of the effect, let us consider two 
cases. In the good-solvent regime, for $\Phi=10$, $L = 250$, we have 
performed simulations with $M=32$ ($N=100$) and $M=64$ ($N=802$). For 
$k = 2 \pi/32$, the smallest momentum which is present in both cases, 
we obtain $S({\bf k};M) = 0.01319(4)$ and $0.01316(2)$ for $M=32,64$,
respectively. Clearly, the systematic error on the estimate obtained 
by using $M=64$ (the only one we use) is negligible. In the opposite 
limit, consider $z=z_1$, $\Phi = 15$, $L=2000$. For $k = 2\pi/64$, we 
obtain $S({\bf k};M)  = 0.09831(6)$, 0.09853(8) for $M=64,128$, respectively. 
Again, Eq.~(\ref{Sk_FSS}) indicates that finite-volume corrections 
on the largest-lattice result are significantly smaller than statistical
errors.  Hence, to obtain the TPM results it is enough to extrapolate 
the finite-$L$ data at the same value of $\Phi$ by using 
Eqs.~(\ref{extr-R}) or (\ref{extr-V}).

In the simulation we estimate $S({\mathbf k})$ for two different
wave vectors: ${\mathbf k}_1 = (2 \pi/M,0,0)$ and
${\mathbf k}_2 = (4 \pi/M,0,0)$, where $M$ is the linear size of the
cubic box. Then, we define \cite{Pelissetto-08}
\begin{equation}
 K_{\rm est} = {\hat{k}_2^2 - \hat{k}_1^2 \over 
        \hat{k}_2^2 S({\mathbf k}_1) - \hat{k}_1^2 S({\mathbf k}_2) },
\label{Kest-def}
\end{equation}
where $\hat{k} = 2 \sin (k/2)$, $k_1 = 2\pi/M$, and $k_2 = 4 \pi/M$.
As discussed in Ref.~\onlinecite{Pelissetto-08}, 
the estimator $K_{\rm est}$ of the inverse compressibility 
converges to $K$ with corrections of order $M^{-4}$, i.e.
\begin{equation}
K_{\rm est}(L,M) = K(L) + {a(L)\over M^4},
\end{equation}
where we have only included the leading correction. 
It turns out that this term is not negligible, hence an additional
extrapolation is needed to obtain $K(L)$.
Of course, $K_{\rm est}$ also depends on $\Phi$ and $z$, but since 
they do not play any role here---we work at fixed density and $z$---
they are omitted. To go further, we must specify the $L$ dependence of 
$a(L)$. Since $a(L)$ corresponds dimensionally to a fourth power 
of a length, we expect on general grounds and verify numerically in a few 
cases that $a(L)\sim \hat{R}_g^4 \sim L^2$,
with corrections that decay as $L^{-1/2}$:
$a(L) = L^2 (b + c L^{-1/2})$ . Since the infinite-volume 
$K(L)$ should behave as $R(L,z)$ in Eq.~(\ref{extr-R}), we end up with
the asymptotic expansion
\begin{equation}
K_{\rm est}(L,M) = K^* + {k\over \sqrt{L}} + 
                  {L^2\over M^4} \left(b + {c\over \sqrt{L}}\right),
\label{K-extr}
\end{equation}
where $K^* = K^*(z,\Phi)$ is the TPM result. For each $\Phi$ and $z$ 
we have therefore fitted our simulation results to Eq.~(\ref{K-extr}),
keeping $K^*$, $k$, $b$, and $c$ as free parameters,
in order to obtain the TPM prediction $K^*(z,\Phi)$. 

\begin{figure}[tbp]
\begin{center}
\begin{tabular}{c}
\epsfig{file=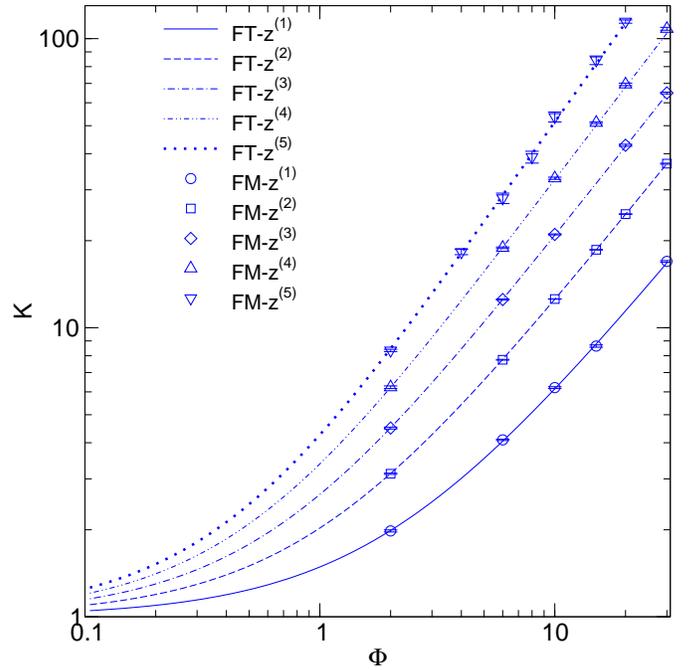,angle=0,width=9truecm} \hspace{0.5truecm} \\
\end{tabular}
\end{center}
\caption{TPM estimates of $K$ for different values of $z$,
see Table~\ref{tab-z}. We report full-monomer (points, FM) and 
field-theory (lines, FT) results. 
}
\label{fig:K-FD}
\end{figure}

It is interesting to compare the numerical results with the 
field-theoretical (FT) predictions of Ref.~\onlinecite{Schaefer-99} 
(the relevant formulae are summarized in App.~\ref{App.B}).
In all cases we observe very good agreement, see
Fig.~\ref{fig:K-FD}. For $z = z^{(1)}$,
$z^{(2)}$, and $z^{(3)}$, in the quite large range of densities
we consider, $\Phi \le 30$, the relative differences are always less
than 1\%. For the two largest values of $z$, differences are somewhat
larger but remain below 4\%. Note that similar conclusions were reached 
in Ref.~\onlinecite{Pelissetto-08} for the good-solvent case.
Hence, the parametrization of Ref.~\onlinecite{Schaefer-99} of the 
FT results appears to be 
quite accurate in the whole crossover regime. 

For $\Phi\to \infty$ and any finite $z$, 
$K(z,\Phi)$ converges\cite{CMP-08} 
to $K_{\rm as}(z,\Phi) \equiv k_{FM}(z)\Phi$ with 
$k_{FM}(z) = 6 \sqrt{\pi} z \alpha_g^{-3}(z)$, where $\alpha_g(z)$ 
is the swelling factor of the zero-density radius of gyration, computed 
explicitly in Ref.~\onlinecite{CMP-08}. For $z = z^{(1)},\ldots,z^{(5)}$
we obtain $k_{FM}(z) = 0.543, 1.27, 2.32, 4.07, 8.40$, respectively.
For each value of $z$,
we can compare our result for the largest value of $\Phi$ with this asymptotic
prediction. For $\Phi = 30$, we obtain from simulation
$K(z,\Phi) = 16.73(6)$, 37.1(2), and 64.8(3), to be compared with 
$K_{\rm as}(z,\Phi) = 16.3$, 38.1, and 69.6, for $z= z^{(1)}$, 
$z^{(2)}$, and $z^{(3)}$, respectively. 
Clearly, for such large value of $\Phi$ the 
asymptotic formula holds approximately only for $z \lesssim z^{(2)}$. 
For larger values of $z$, the linear behavior sets in for values 
of $\Phi$ that are significantly larger than 30.

Finally, we should mention that Ref.~\onlinecite{CMP-08} also gave a 
prediction for $Z(z,\Phi)$ [Eq.~(4.24) of Ref.~\onlinecite{CMP-08}] 
valid for $z \to \infty$ 
and $\Phi\lesssim \Phi_{\rm max}(z)\sim z^{2.53}$. Requiring an error of 
at most 5\%, the asymptotic expansion was found to be predictive for 
$z\gtrsim 2$, $\Phi \lesssim (z/2)^{2.53}$. The first condition implies 
that we can only consider $z = z^{(5)}$, while the second condition 
implies that $\Phi$ should be relatively small, $\Phi\lesssim 2$.
In this density range, by using the equation of state reported in the 
supplementary material,\cite{suppl}
we find that the asymptotic formula reproduces 
$Z(z^{(5)},\Phi)$ with an error of at most 4\%, confirming the 
correctness of the numerical estimates of Ref.~\onlinecite{CMP-08}.

\subsection{Comparison in the semidilute regime} \label{sec4.3}

\begin{figure}[tbp]
\begin{center}
\begin{tabular}{c}
\epsfig{file=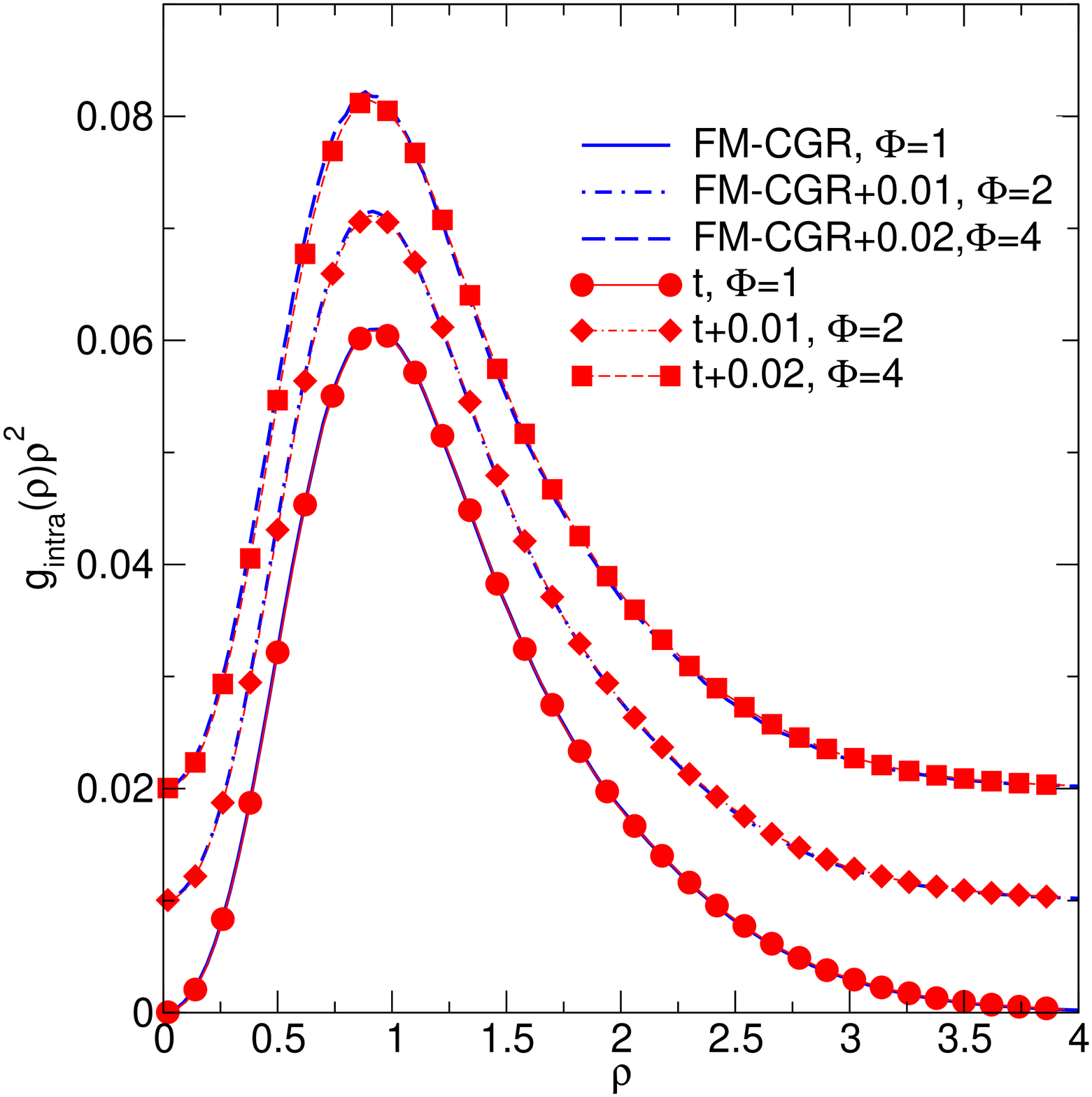,angle=0,width=9truecm} \hspace{0.5truecm} \\
\epsfig{file=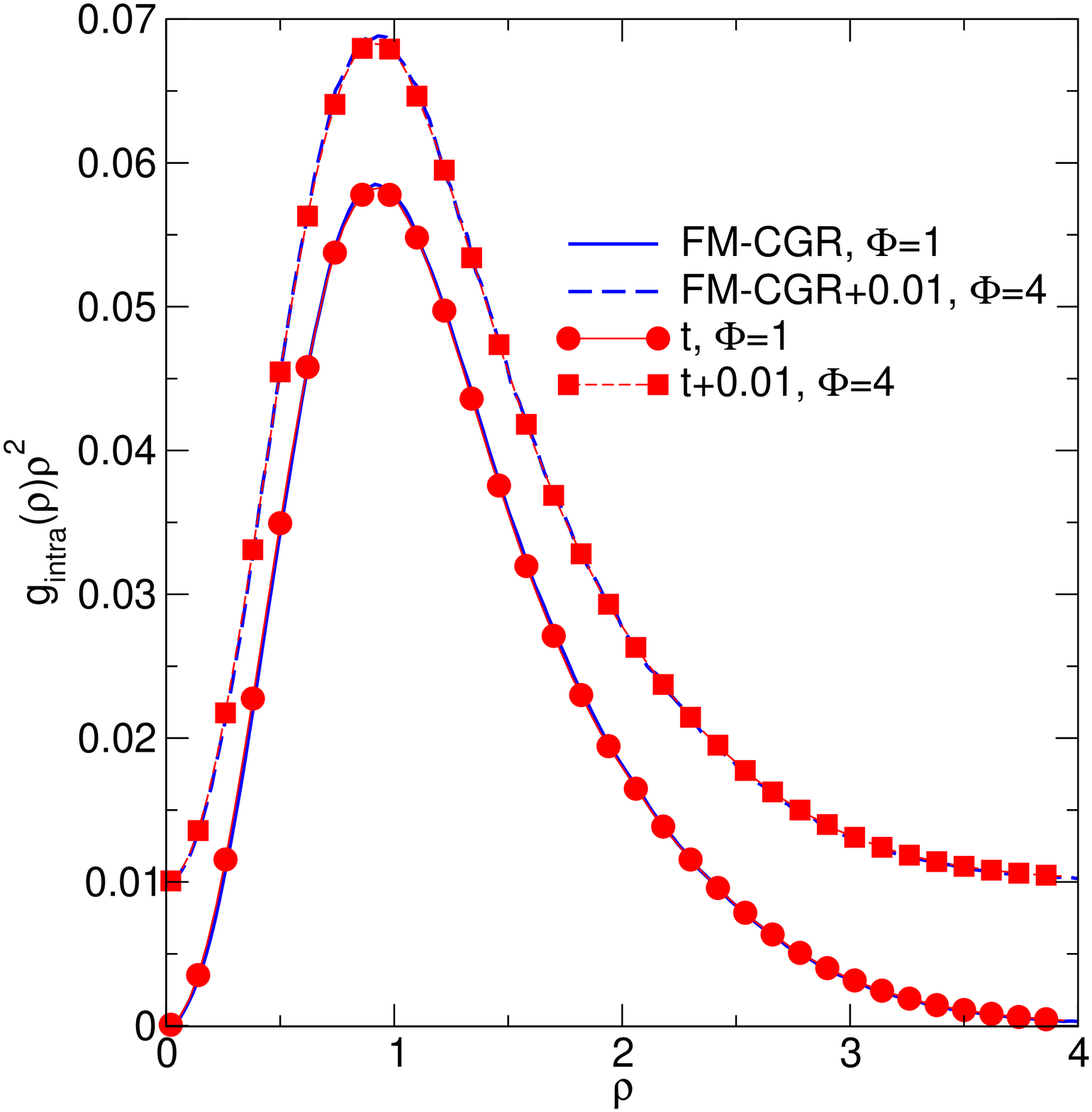,angle=0,width=9truecm} \hspace{0.5truecm} \\
\end{tabular}
\end{center}
\caption{Intramolecular distribution function as a function of 
$\rho = r/\hat{R}_g$, for $z = z^{(3)}$ (top)
and $z = z^{(1)}$ (bottom). We report full-monomer (FM-CGR)
and tetramer (t) results. 
}
\label{fig:intra-FD}
\end{figure}

\begin{figure}[tbp]
\begin{center}
\begin{tabular}{c}
\epsfig{file=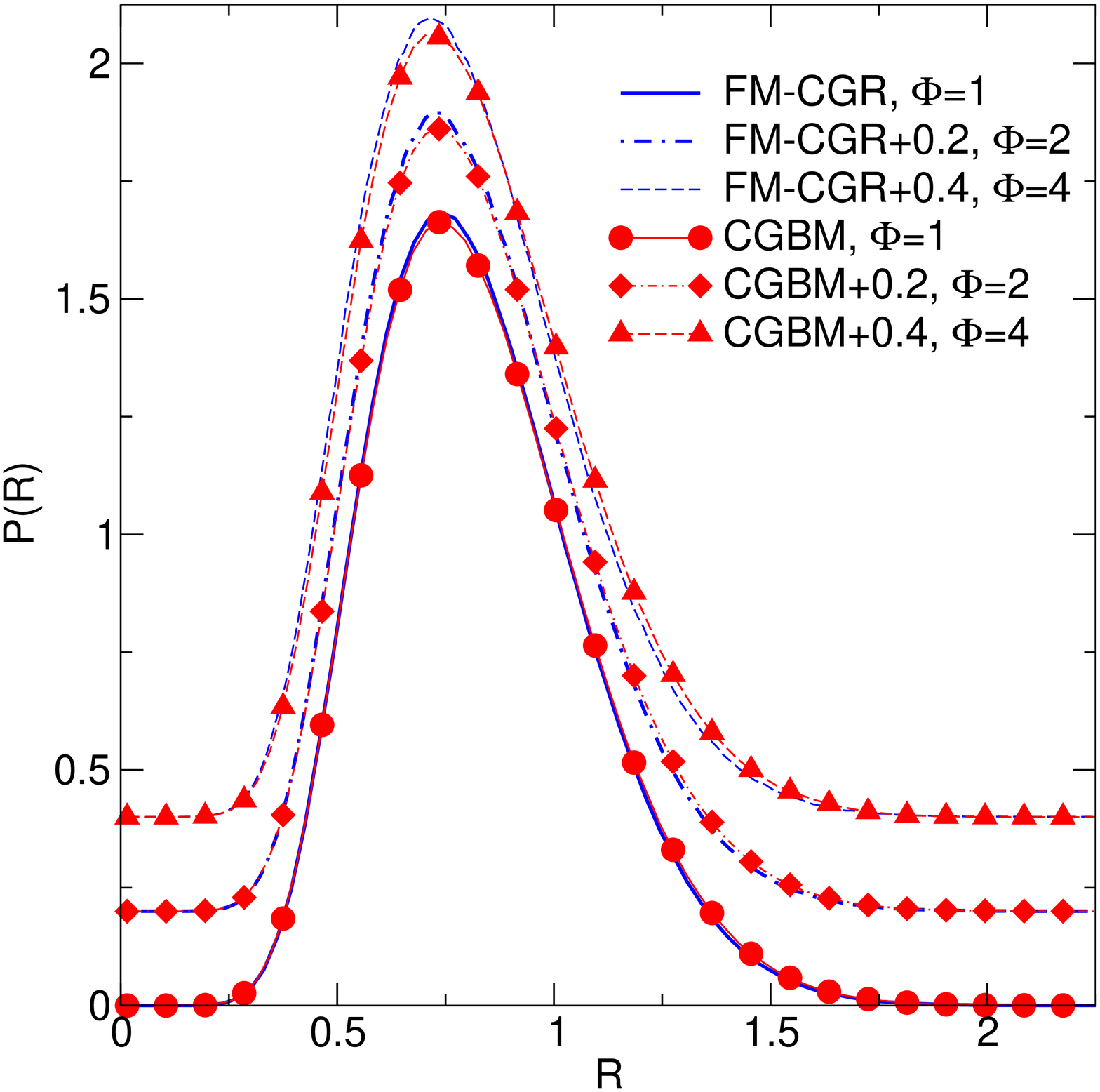,angle=0,width=9truecm} \hspace{0.5truecm} \\
\epsfig{file=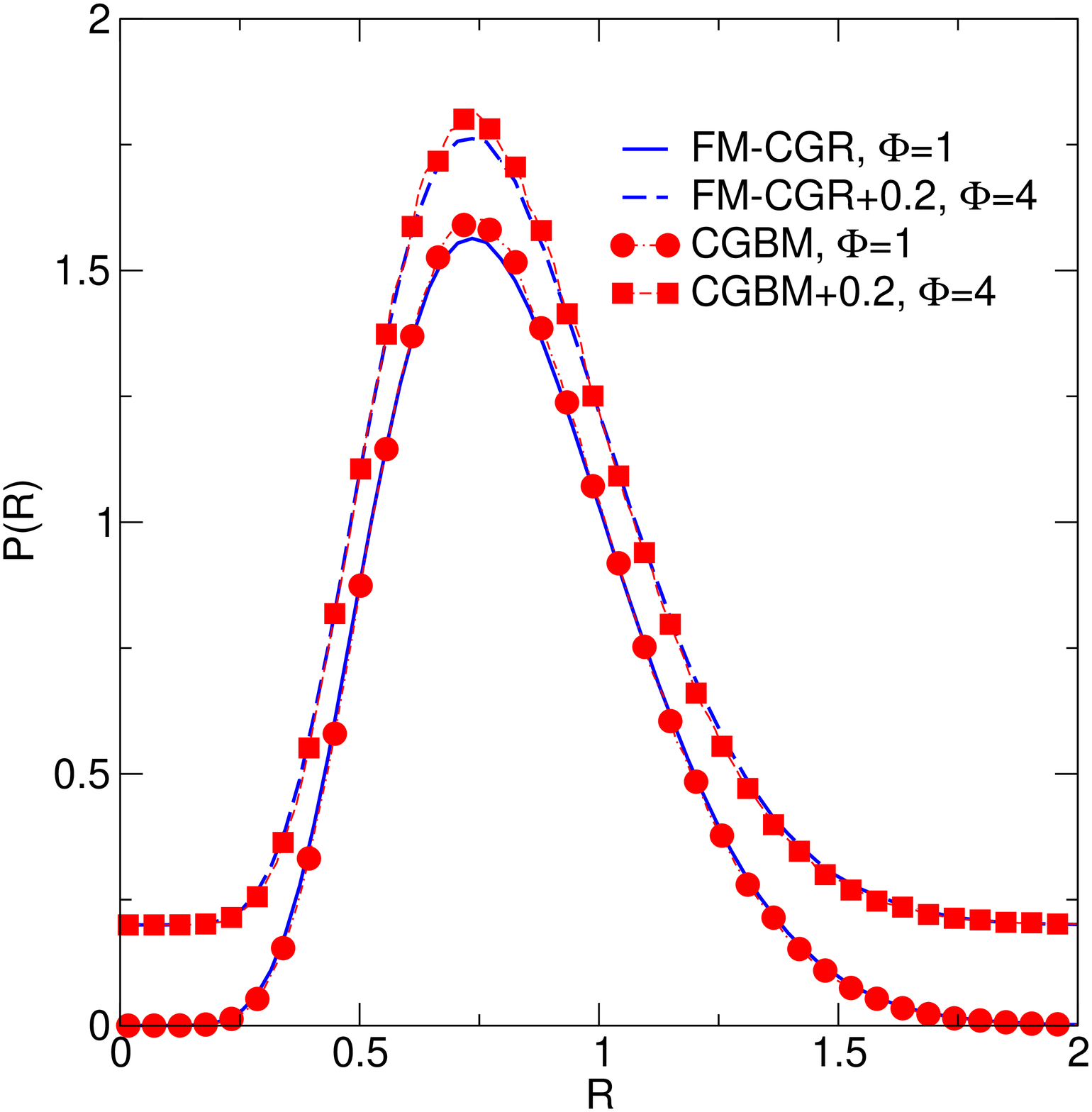,angle=0,width=9truecm} \hspace{0.5truecm} \\
\end{tabular}
\end{center}
\caption{Distribution of $R = R_{g,b}(\Phi)/\langle \hat{R}_g\rangle$
for $z = z^{(3)}$ (top) 
and $z = z^{(1)}$ (bottom). We report full-monomer (FM-CGR)
and tetramer (t) results. 
}
\label{fig:Rgbdist-FD}
\end{figure}

Let us now compare the CGM predictions with the polymer CGR results 
at finite density.
Let us begin by comparing the intramolecular distribution function.
Results for $z=z^{(1)}$ and $z=z^{(3)}$ are reported in 
Fig.~\ref{fig:intra-FD}. In all cases the agreement is excellent, even for 
$\Phi = 4$. Similar conclusions are reached for the 
finite-density distribution of $R_{g,b}$, see Fig.~\ref{fig:Rgbdist-FD}.
Clearly, the tetramer model correctly reproduces the large-scale structure
of the polymer even in the presence of significant polymer-polymer 
overlap. Note that this is not the case in the good-solvent regime.
\cite{DPP-12-Soft} In that case, for $\Phi\approx 4$ the tetramer is more 
swollen than the polymer: the probability for two blobs 
to be at a given distance $\rho\lesssim 1$ is significantly smaller
in the tetramer than in the polymer. Analogously, $R_{g,b}$ is typically
larger for the tetramer than for the polymer. 

\begin{figure}[tbp]
\begin{center}
\begin{tabular}{c}
\epsfig{file=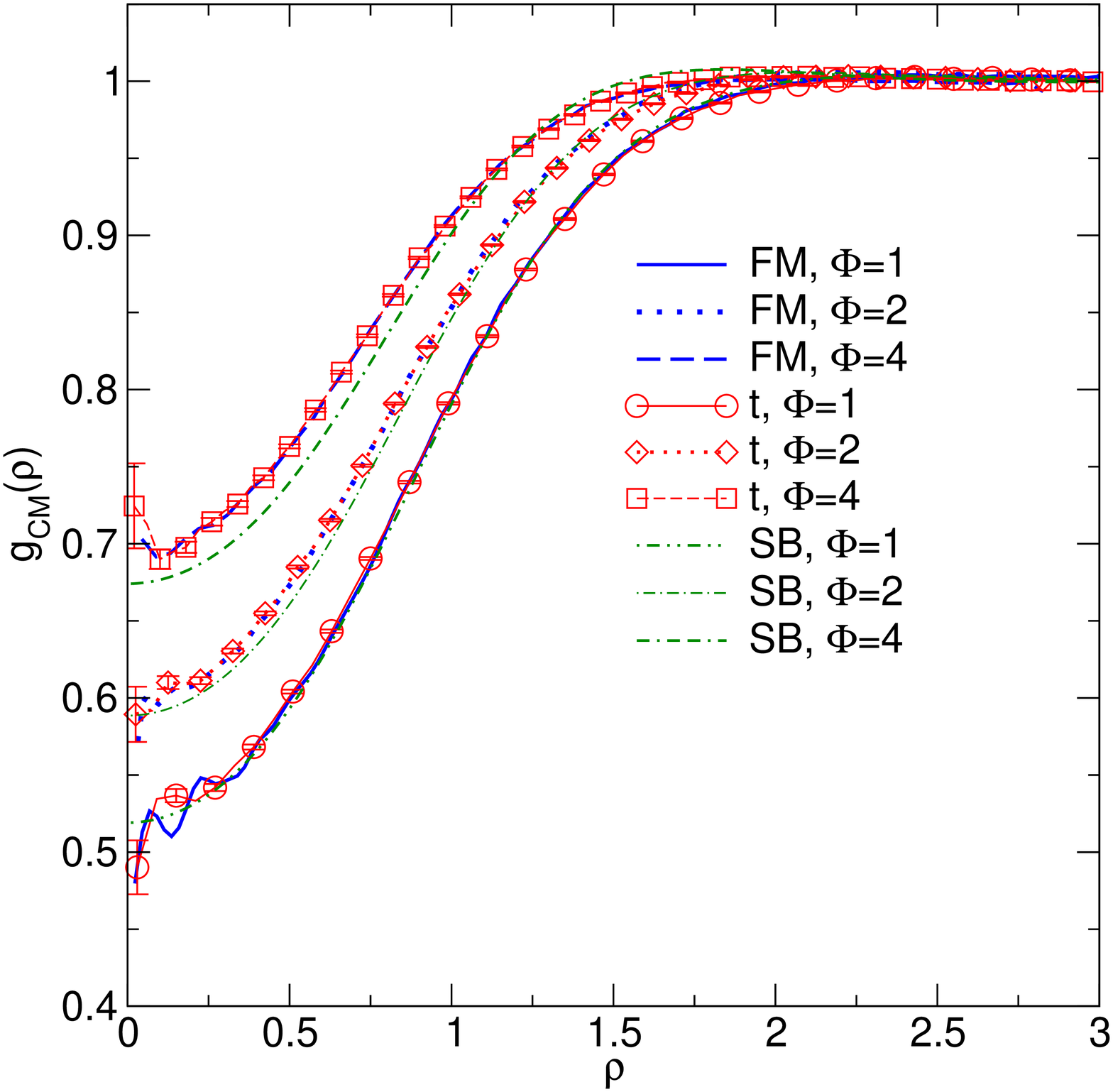,angle=0,width=9truecm} \hspace{0.5truecm} \\
\epsfig{file=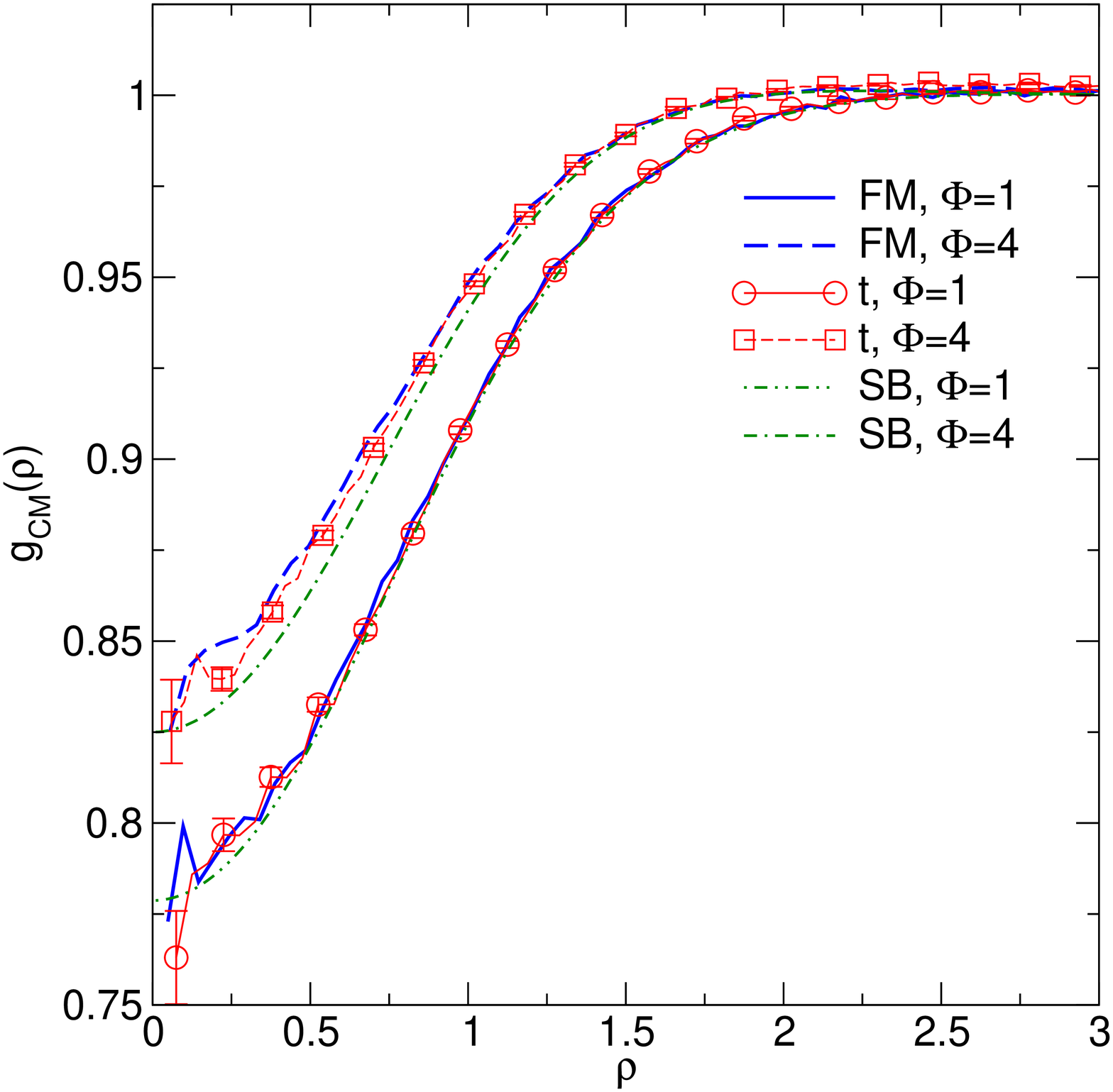,angle=0,width=9truecm} \hspace{0.5truecm} \\
\end{tabular}
\end{center}
\caption{Intermolecular center-of-mass pair distribution function
as a function of $\rho=r/\hat{R}_g$, for $z = z^{(3)}$ (top)
and $z = z^{(1)}$ (bottom).
We report full-monomer (FM), single-blob (SB) and tetramer (t) results. }
\label{fig:CMCM-FD}
\end{figure}

Let us now compare the intermolecular structure. In Fig.~\ref{fig:CMCM-FD}
we report the center-of-mass distribution function $g_{CM}(r)$ for the 
tetramer, the SB model, and from FM simulations. For $z=z^{(1)}$
both the tetramer and the SB model correctly reproduce the polymer 
structure, even for $\Phi = 4$. Clearly, many-body interactions are weak,
hence do not influence significantly the polymer behavior, even in the 
presence of significant overlap among the polymers. For $z=z^{(3)}$ the 
SB model reproduces $g_{CM}(r)$ for $\Phi = 1$, 
while discrepancies are observed for $\Phi = 4$. On the other hand, 
the tetramer model is quite accurate, even for $\Phi = 4$. Many-body
interactions are relevant, but are reasonably taken into account by the 
tetramer model.

\begin{figure}[tbp]
\begin{center}
\begin{tabular}{c}
\epsfig{file=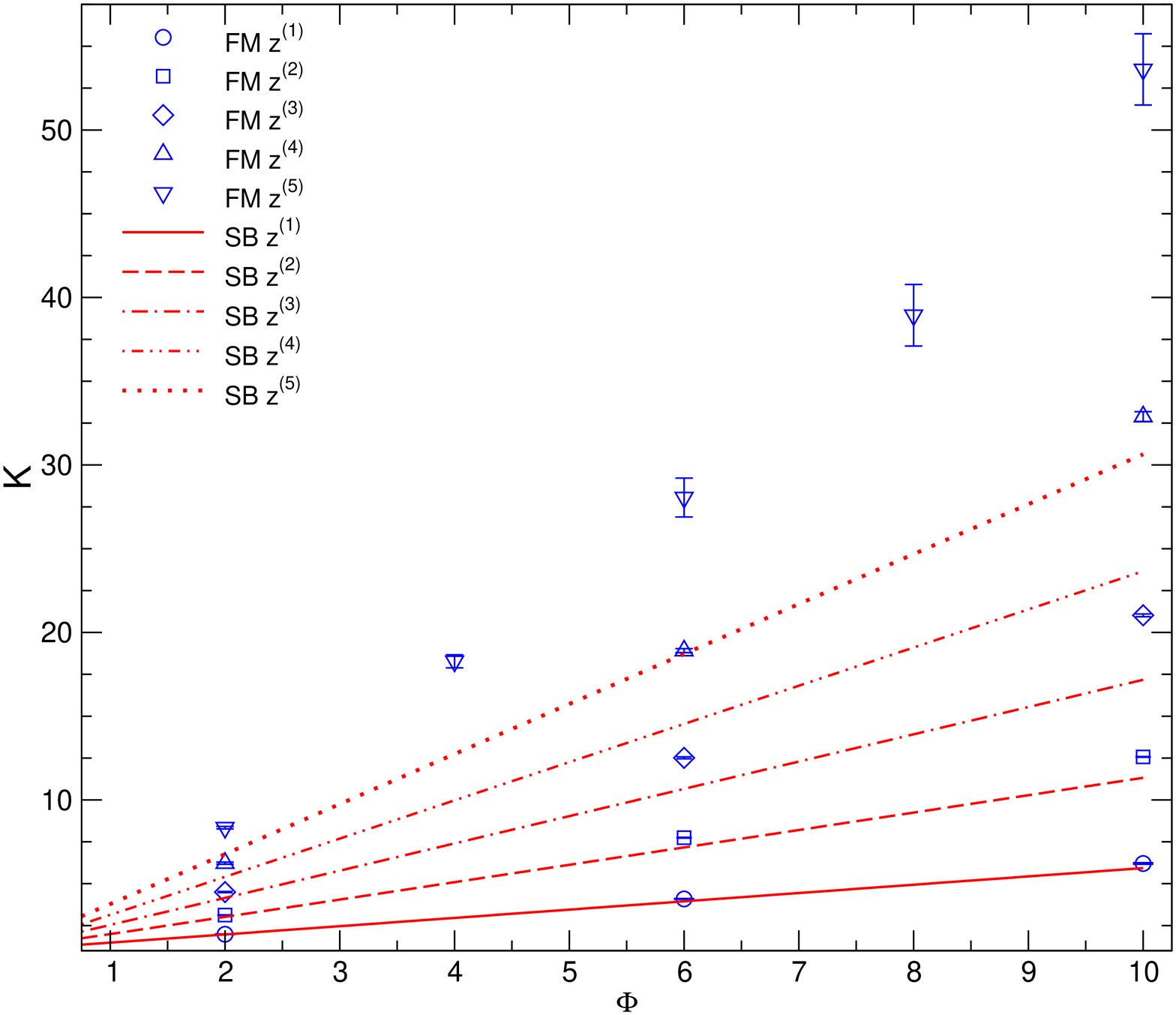,angle=0,width=9truecm} \hspace{0.5truecm} \\
\epsfig{file=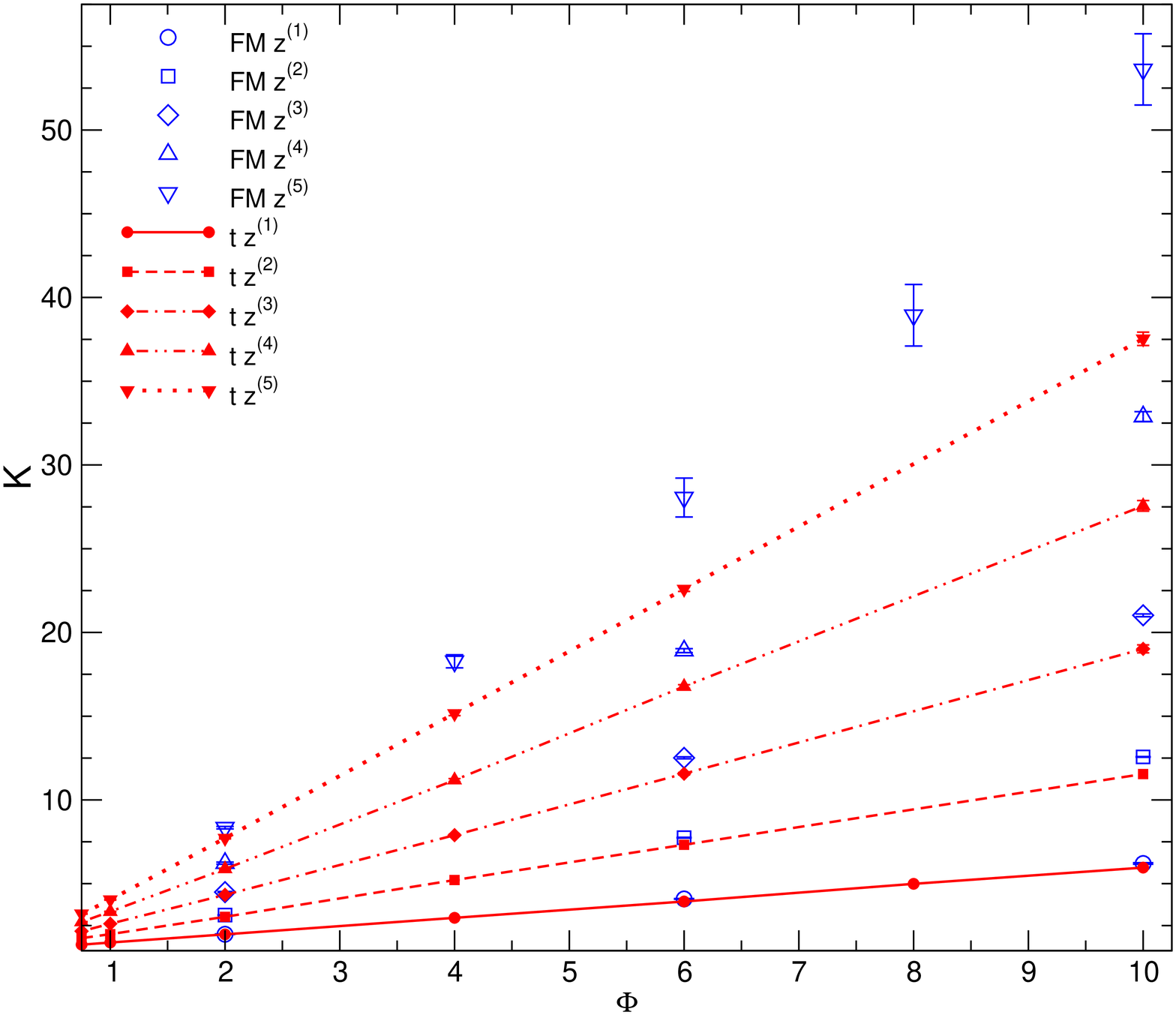,angle=0,width=9truecm} \hspace{0.5truecm} \\
\end{tabular}
\end{center}
\caption{Inverse compressibility $K$ as predicted by 
full-monomer (FM) simulations, the single-blob (SB) 
model (top), and the tetramer (t) model (bottom).
}
\label{fig:EOS-comparison}
\end{figure}

Let us finally, discuss the thermodynamics, comparing the estimates 
of $K(z,\Phi)$ for polymers with those obtained by using the 
SB and the tetramer model, see Fig.~\ref{fig:EOS-comparison} 
(see supplementary material\cite{suppl} for tables of data).
For $z = z^{(1)}$ tetramer results are on top of the FM results, 
even for $\Phi = 10$: at this density we obtain $K = 5.96(4)$ (tetramer), 
$K = 6.15(3)$ (FM). For $z=z^{(2)}$ discrepancies are small 
(the relative difference is 5\% for $\Phi = 6$ and 8\% for $\Phi=10$) 
and so are they for $z = z^{(3)}$ (9\% for $\Phi = 10$). 
For $z = z^{(4)}$ and $z^{(5)}$  we observe larger differences,
the relative difference being larger than 10\% for $\Phi\gtrsim 5$ and 3,
respectively. Thermodynamic results are completely consistent with 
the structural ones. For $z\lesssim z^{(3)}$, i.e. for systems such that 
$A_2\lesssim 3$ (i.e., 
$A_2/A_{2,GS} \lesssim 0.55$, where $A_{2,GS} = 5.50$ is the 
good-solvent value), the tetramer model reasonably works also deep in the 
semidilute regime. On the other hand, for larger values of $z$, many-blob
interactions begin to play a role. Hence, in this region,
the model is only predictive for $\Phi$ not too large, i.e. for densities
such that blob-blob overlaps are rare. 

It is also interesting to compare the thermodynamic predictions for the 
SB model. For $z=z^{(1)}$ and $z=z^{(2)}$ the SB model gives estimates of $K$
that are close to the tetramer and FM ones. Differences are observed 
for $z\gtrsim z^{(3)}$: in this regime the SB model significantly 
underestimates $K$ as long as $\Phi\gtrsim 1$.

\subsection{Comparison for $\Phi\to\infty$ and random-phase approximation}
\label{sec4.4}

\begin{table}[tbp!]
\caption{Large-$\Phi$ coefficients $k_{SB}(z)$, $k_4(z)$, and $k_{FM}(z)$
for five intermediate values of $z$, see Table~\ref{tab-z}, and for the 
good-solvent case ($z=\infty$). For $n\ge 4$, $k_n(z=\infty) = 3.204
n^{2-3\nu}$. 
}
\label{tab:kz-1}
\begin{center}
\begin{tabular}{cccc}
\hline
\hline
$z$ & $k_{SB}(z)$ & $k_4(z)$   & $k_{FM}(z)$   \\
\hline
$z^{(1)}$   &	0.4956   & 0.501 & 0.543  \\
$z^{(2)}$   & 	1.0352   & 1.083 & 1.270  \\
$z^{(3)}$   & 	1.6271   & 1.778 & 2.320  \\
$z^{(4)}$   &	2.2750   & 2.643 & 4.072  \\
$z^{(5)}$   &	2.9731   & 3.643 & 8.504  \\
$\infty$    &   3.4163   & 4.584 & ---    \\
\hline\hline
\end{tabular}
\end{center}
\end{table}

We wish now to compare the thermodynamic behavior for large values of 
$\Phi$. For the polymer system, we have 
$K(z,\Phi) \approx K_{\rm as}(z,\Phi) = k_{FM}(z)\Phi$, where 
$k_{FM}(z)$ is reported in Sec.~\ref{sec4.2}. We now compute the large-$\Phi$
behavior of $K(z,\Phi)$ for the CG models, using the fact that 
the random-phase approximation (RPA)\cite{HMD-06} is exact at large density
for systems with soft potentials. For the SB model, we start from the 
virial pressure
\begin{equation}
Z_{SB}(z;\Phi) = 1 - {\Phi\over2} \int_0^{\infty} 
      {\partial \beta V_{SB}(\rho;z)\over \partial \rho} g_{CM}(\rho;z,\Phi)\,
       \rho^3d\rho.
\end{equation}
In the RPA, we set $g_{CM}(r;z,\Phi)=1$, a property which is rigorously
true for $\Phi\to\infty$. Integrating by parts we end up with the 
usual RPA expression\cite{HMD-06}
\begin{eqnarray}
&& Z_{SB}(z,\Phi) = 1 + {1\over2}\Phi k_{SB}(z) 
\nonumber \\
&&   k_{SB}(z) = 3 \int_0^\infty \beta V_{SB}(\rho;z)\, \rho^2 d\rho,
\label{RPA}
\end{eqnarray}
from which we obtain $K_{SB}(z,\Phi) = 1 + \Phi k_{SB}(z)$. 
The function $k_{SB}(z)$ is reported in the 
supplementary material\cite{suppl} and in Table~\ref{tab:kz-1} for some 
specific values of $z$.
The RPA expression (\ref{RPA}) reproduces the SB compressibility factor 
at the 1\% level for $z = z^{(1)}$ for all
values of $\Phi$. For larger values of $z$ it provides a very good
approximation for
$\Phi\gtrsim 3$ (deviations are at most 2\% for $\Phi=3$ and 
are less than 1\% for $\Phi\ge 8$). For smaller densities deviations are 
larger, due to the fact that the small-density behavior is not correctly
reproduced (compare $k_{SB}(z)$ with the estimates of $A_2$ reported in 
Table~\ref{tab-z}).

The RPA expression (\ref{RPA}) can be extended to the tetramer model.
Starting from the expression of the pressure in terms of the 
atomic virial and assuming the absence of blob-blob correlations 
among blobs belonging to different tetramers,
we obtain for a CGM with $n$ blobs
(see also Ref.~\onlinecite{PCH-07})
\begin{equation}
Z(z,\Phi) = 1 + {1\over2} \Phi k_n(z) \qquad
   k_n(z) = 3 n^2 \int_0^\infty \beta W(\rho;z)\, \rho^2 d\rho.
\label{Zlarge-n-RPA}
\end{equation}
where $W(\rho;z)$ is the intermolecular potential.
Estimates of $k_4(z)$ are reported in Table~\ref{tab:kz-1}. 
In the good-solvent case, using the transferability assumption, the 
integral should scale as $n^{-3\nu}$, so that we obtain
$k_n(z=\infty) = 3.204 n^{2-3\nu}$ (the prefactor has been fixed 
by using the explicit rescaling factor reported in
Ref.~\onlinecite{DPP-12-JCP}). 

Knowledge of the constants $k(z)$ allows us to compute the deviations 
between CGM and FM results for $\Phi\to \infty$ without the need 
of simulations, since
\begin{equation}
{K_n(z,\Phi)\over K_{FM}(z,\Phi)} \approx {k_n(z)\over k_{FM}(z)}
\label{discrepancy-RPA}
\end{equation}
for $\Phi\to\infty$.
For $z = z^{(1)}$, $k_{SB}(z)$ and $k_4(z)$ are close to 
$k_{FM}(z)$: the SB $K(z,\Phi)$ (or equivalently $Z(z,\Phi)$)
differs from the corresponding polymer quantity by 9\% for large $\Phi$. 
For such a small value of $z$, there is little advantage in using the 
tetramer model if one is only interested in the thermodynamics: 
the difference decreases only to 8\%. For $z = z^{(2)}$
differences are larger: 18\% and 15\% for the SB and tetramer model, 
respectively. For larger values of $z$, instead, it is quite clear that 
both the SB and the tetramer model grossly underestimate the correct
polymer pressure deep in the semidilute regime.


\section{Transferability in the number of blobs} \label{sec5}

\subsection{Transferability: general considerations} \label{sec5.1}

In the previous section we have determined a CG tetramer model 
appropriate to describe the $\theta$-to-good-solvent crossover. 
It was found that the predictions of the tetramer model are accurate 
up to a value of the density $\tilde{\Phi}(z)$,
which increases when $z$ decreases towards the $\theta$ region.   
Above $\tilde{\Phi}(z)$ the tetramer model is observed to deviate from the 
FM results and to approach the RPA predictions at large $\Phi$.
We wish now to 
extend the CG model to a larger number $n$ of blobs per chain,
which is the key ingredient to increase $\tilde{\Phi}$. 
In the good-solvent case,
the basic transferability assumption is that the tetramer potentials, 
expressed in terms of the blob radius of gyration $\hat{r}_g$, 
provide an accurate CGM for any number of blobs. 
For $z=\infty$, the approach is therefore
the following. First, we express any tetramer bonding potential 
$V_{ij}(\rho;n=4)$,
$\rho = r/\hat{R}_g$,
in terms of $\sigma = r/\hat{r}_g$:
\begin{equation}
 \hat{V}_{ij}(\sigma;n=4) = V_{ij}(\sigma{\cal R}_4),
\end{equation}
where ${\cal R}_4 = \hat{r}_g/\hat{R}_g$ for the tetramer. Then, the potentials
for the $n$-blob model are defined by
\begin{eqnarray}
&&\hat{V}_{12}(\sigma;n) = \hat{V}_{n-1,n}(\sigma;n) = \hat{V}_{12}(\sigma;4), 
\nonumber \\
&&\hat{V}_{ij}(\sigma;n) = \hat{V}_{23}(\sigma;4) \qquad\qquad
  |i-j|=1, i\not=1,n-1, \nonumber \\
&&\hat{V}_{ij}(\sigma;n) = \hat{V}_{13}(\sigma;4) \qquad\qquad
  |i-j|=2, \nonumber \\
&&\hat{V}_{ij}(\sigma;n) = \hat{V}_{14}(\sigma;4) \qquad\qquad
  |i-j|\ge 3. 
\label{corrispondenze-pot}
\end{eqnarray}
Note, that blobs $i$ and $j$ with $|i-j|>3$ always interact 
with potential $\hat{V}_{14}(\sigma;4)$
to guarantee the local self-repulsion, which is necessary to obtain
a good-solvent CGM.
The angular potentials are instead unchanged when increasing 
the number of blobs per chain.  If 
$\hat{R}_g$ is used as reference length scale, i.e. we consider 
$V_{ij}(\rho;n)$ with $\rho = r/\hat{R}_g$, these relations imply the 
rescalings $V(\rho;n) = V(\rho {\cal R}_4/{\cal R}_n;4)$.\cite{DPP-12-JCP}

To justify the previous relations, we should first note the 
dual interpretation of the coarse-graining procedure. Up to now,
we have considered two blob models with $n_1$ and $n_2$ blobs each
as providing two different representations (with different
resolutions) of the same polymer chain. However, since polymer chains
are fractals, hence scale invariant, for large degree of polymerization 
$L$, the multiblob model can be given a different interpretation.
We assume now that the number $m$ of monomers belonging to a blob is fixed,
so that CGMs with $n_1$ and $n_2$ blobs are CGRs with the {\em same} 
resolution of polymer chains of different lengths, $L_1 = n_1 m$ and 
$L_2 = n_2 m$, respectively. Within this dual interpretation it is easy
to justify transferability.
First, we note that, to a very good approximation, the size of the blob 
depends only on $m$ and not on the length of the chain.
The latter nontrivial property was verified for
$n\ge 4$ in App.~A of Ref.~\onlinecite{DPP-12-Soft}. Indeed,
since $\hat{R}_g \sim L^\nu$ and 
$\hat{r}_g/\hat{R}_g \sim n^{-\nu}$ with
good precision for $n\ge 4$, we have $\hat{r}_g \sim m^\nu$, which
depends only on $m$ and not on $L$.
Therefore, $\hat{r}_g$ is the same for the two CGMs. 
Second, we 
assume that the interactions have little dependence on the chemical 
distance between the blobs, except for the case in which the 
blobs are very close along the chain, and are insensitive to the 
length of the CGM. If this holds, potentials
are the same for the two CGMs, i.e. they can be {\em transferred}
from one model to the other one. 
If we wish to use this result within the original 
CG interpretation, we should simply note that potentials are 
invariant, if $\hat{r}_g$ is the basic length scale. If $\hat{R}_g$ is used 
instead, a rescaling of the length scale should be performed.

Let us now consider the case of polymer chains in the thermal crossover region. 
Let us consideri, at the same temperature, two chemically identical chains 
of length $L_1$ and $L_2$, respectively, and their 
CGRs in terms of $n_1$ and $n_2$ blobs, each blob
consisting of the same number $m$ of monomers. If we assume that 
(i) the size of the blob is the same for the two chains (this hypothesis 
looks reasonable, but we shall show below that it is only a rough
approximation)
and that (ii) interactions are independent on chemical distance
except for very close blobs along the chain
(again this looks quite plausible),
we expect the potentials to be approximately 
transferable without any change as in the good-solvent
case. However, since $z = (T-T_\theta) L^{1/2}$, the 
two chains correspond to two different crossover parameters $z_1$ and $z_2$. 
Therefore, the CGM with $n>4$ blobs obtained by using 
the  tetramer potentials at $z$ is a CGM for a system at $z'> z$. 
In the renormalization-group language, $z$ flows towards 
the stable good-solvent fixed point as the number of blobs increases.
If hypotheses (i) and (ii) above were both approximately correct, 
we could simply estimate $z' = z (n/4)^{1/2}$.
However,
since $z'\not= z$, we expect $\hat{r}_g$ to differ in the two cases, 
hence hypothesis (i) does not hold,
and therefore the relation between $z$ and $z'$ is more complex.

It is interesting to revisit this argument within the usual
interpretation of the coarse-graining procedure, in which an increase of $n$
corresponds to an increase of the resolution of the polymer CGR.
In this case $z$ is fixed. The previous argument implies that 
potentials should become increasingly softer as the resolution increases,
i.e., the effective $z$ decreases with increasing $n$.
This result is completely 
consistent with the general argument of deGennes,\cite{deGennes-79}
who noted that, 
close to the $\theta$ point, polymers show two different spatial 
regimes. If we indicate with $R_t$ the thermal blob size,
\cite{deGennes-79} 
blobs such that  $\hat{r}_g\ll R_t$ behave as ideal chains, while 
excluded-volume effects dominate for $\hat{r}_g\gg R_t$. 
For $\hat{r}_g\sim R_t$, which is the relevant case here, 
blobs behave in an intermediate way, excluded-volume effects 
becoming increasingly less relevant --- hence the effective $z$ decreases ---
as $\hat{r}_g$ decreases, i.e. when the resolution (number of blobs $n$) 
increases.

Given the difficulties presented above, we have developed a 
transferability procedure which works at a formal level, 
without any explicit reference to the underlying polymer system. 
The idea is the following. Consider the tetramer set of potentials 
$\{V_4\}$ at a given value of $z$, $z=z_4$, and a system of $n>4$ blobs
interacting with these potentials. The question we will ask 
is: can this $n$-blob system be seen as a CGR of a polymer chain
at a different value of $z$, say $z_n$? As we shall show below, the answer
is positive. We will compute $z_n$ and we will relate the size of the 
new CGR, i.e. $\hat{R}_{g,b}$, to the size of the corresponding underlying 
polymer chain. In terms of the underlying model, the tetramer potentials 
appropriate to describe a polymer chain at temperature $T$ can be transferred
without changes 
to an $n$ blob CGR of the same polymer chain, 
but at a different temperature $T'>T$. In the renormalization-group (RG)
language, we are considering a RG transformation at fixed {\em bare} 
parameters, hence we must consider the temperature RG flow 
towards the good-solvent fixed point $T=\infty$. 

\subsection{Definition of the higher-resolution models} \label{sec5.2}

\begin{table}[tb]
\caption{The transferability mapping: for each $z$ ($z_4$)
in the tetramer model
($A_{2-4}$ is the corresponding second-virial combination)
we report the value of $z$ ($z_{10}$), the 
corresponding second-virial combination $A_{2-10}$, and the rescaling
factor $\lambda$ [see Eq.~(\ref{def-lambda})] for the decamer model.  
}
\label{table:transferability}
\begin {tabular}{cclll}
\hline\hline
 \multicolumn{2}{c}{$n=4$} & 
 \multicolumn{3}{c}{$n=10$} \\
$z_4$ &  $A_{2-4}$ & 
\multicolumn{1}{c}{$z_{10}$} & $A_{2-10}$  &
\multicolumn{1}{c}{$\lambda$} \\
\hline
$z^{(1)} = 0.056215$ & 0.993 & $z^{(10-1)} = 0.08628$ & 1.374 & 0.9875\\
$z^{(2)} = 0.148726$ & 1.978 & $z^{(10-2)} = 0.22953$ & 2.524 & 0.9833\\
$z^{(3)} = 0.321650$ & 2.962 & $z^{(10-3)} = 0.47018$ & 3.442 & 0.9684\\
$z^{(4)} = 0.728877$ & 3.943 & $z^{(10-4)} = 1.00862$ & 4.267 & 0.9604\\
$z^{(5)} = 2.508280$ & 4.915 & $z^{(10-5)} = 3.74117$ & 5.088 & 0.9778\\
\hline\hline
\end{tabular}
\end{table}

Our recipe to transfer the tetramer potentials to $n$-blob CGMs works 
as follows.
First, we introduce two universal functions: We define 
$F_{A,b}(A_2,n) = B_2(z)/[\hat{R}_{g,b}(z,n)]^3$ 
[$B_2$ is the second virial coefficient defined in Eq.~(\ref{eq:virial})],
which is a universal function of $z$ and $n$ or,
equivalently, of $A_2$ and $n$,
and $S_b(A_2,n) = \hat{R}_{g,b}(z,n)/\hat{R}_g(z)$ 
which is also a universal function of $A_2$ and $n$. Of course, the 
two functions are related by
$F_{A,b}(A_2,n) = A_2 S_b(A_2,n)^{-3}$.
These two functions can be computed by FM simulations
and are assumed known in the procedure we shall present. 
For $n=10$, the case we will be interested in, they are reported in 
Appendix~\ref{App.A}.

The procedure is the following:
\begin{itemize}
\item[(i)] 
We consider the $n$-blob model with the same potentials 
used for the tetramer. For the bonding potentials we set $V_{ij}(\rho;n) = 
V_{ab}(\rho;4)$, where $ij$ are related to $ab$ as in 
Eq.~(\ref{corrispondenze-pot}), while angular potentials are unchanged. 
No rescalings are performed at this stage, so that $\rho$ for the 
$n$-blob model should not be identified with $r/\hat{R}_g$.
We will write therefore $\rho = r/R$, where $R$ simply sets the length
scale but is otherwise arbitrary (in the numerical computations we 
set $R=1$).
Then, we compute (for instance,
by Monte Carlo simulations) the 
second virial coefficient $B_{2,MC}$, the radius of gyration
$\hat{R}_{g,b,MC}$, and $A_{2,b,MC} = B_{2,MC}/\hat{R}_{g,b,MC}^3$
of the $n$-blob chain. 
\item[(ii)]
We determine the value of 
$A_2$ appropriate for the $n$-blob model, i.e. $A_2(z_n,n)$,
by solving the equation $F_{A,b}(A_2,n) = A_{2,b,MC}$.
Then, using Eq.~(\ref{A2-to-z}), we determine the $n$-blob corresponding 
$z_n$.
\item[(iii)]
The radius of gyration $\hat{R}_{g}$ of the polymer chain, whose 
CGR is provided by the $n$-blob chain, is given by 
$\hat{R}_{g,MC} = \hat{R}_{g,b,MC}/S_b(A_2,n)$, 
using the value of $A_2$ computed at point (ii).
\item[(iv)]
Once $\hat{R}_{g}$ is known, if we wish to 
express all lengths in terms of the radius of gyration, it is enough 
to redefine the bonding and the intermolecular potentials as 
\begin{equation}
 V(\rho';z_n;n) = 
 V(\rho' \hat{R}_{g,MC}/R;z;4),
\end{equation}
where $\rho' = r/\hat{R}_g(z_n;n)$.
\end{itemize}
With these definitions, we obtain a model with a higher resolution 
at a different value $z_n$, which has 
the correct $A_2$, hence it gives the correct thermodynamics, and 
gives the correct result for $\hat{R}_{g,b}(z;n)/\hat{R}_g(z;n)$, hence it 
is also structurally consistent.


We have applied this strategy starting from the tetramer potentials at 
$z=z^{(1)},\ldots, z^{(5)}$. 
For each of them we have determined a decamer model with 
$n=10$ blobs. The corresponding values of $z_{10}$ and $A_2(z_{10})$ are 
reported in Table \ref{table:transferability}.
To clarify the procedure, let us show how the method works in a specific 
example, applying the transferability procedure to the tetramer model at
$z = z^{(1)} = 0.056215$ 
(Explicit expressions for the
corresponding potentials are reported in the supplementary material
\cite{suppl}).
Consider the decamer model with $n=10$,
using the tetramer potentials (no rescalings are performed). 
For this model we determine numerically
[point (i)]
$B_{2,MC} R^{-3} \approx 5.740$, $\hat{R}_{g,b,MC}/R \approx 1.532$, and
$A_{2,b,MC} \approx 1.597$. Then [point (ii)], we first solve 
the equation
$F_{A,b}(A_2,10) = 1.597$ ($F_{A,b}(A_2,10)$ is given in App.~\ref{App.A}), 
obtaining $A_2 \approx 1.375$, and then the equation
$A_2(z_{10}) = 1.375$ [$A_2(z)$ is given in Eq.~(\ref{A2-to-z})],
obtaining $z_{10} \approx 0.0862$. 
For such value of 
$A_2$, the results of App.~\ref{App.A} give $S_b(A_2,10) \approx 0.951$, hence 
$\hat{R}_{g,MC}/R \approx 1.610$. Hence, for the decamer model, if we set
$\rho' = r/\hat{R}_g$ expressing all lengths in terms of the 
radius of gyration of the underlying polymer chain, 
we should rescale the potentials as 
$V(\rho';10) = V(1.61\rho';4)$.
In the good-solvent case, the rescaling is equal to the 
ratio ${\cal R}_4/{\cal R}_n = \hat{r}_{g,4}/\hat{r}_{g,n}$. 
In the crossover region, this relation does not 
hold, because of the flow of $z$. Hence we define 
\begin{equation}
\lambda = {\hat{r}_{g}(z_4,4)\over \hat{r}_{g}(z_4,n)} {R\over \hat{R}_{g,MC}},
\label{def-lambda}
\end{equation}
which encodes how much of the length rescaling is due to the 
change of the parameter $z$. At the renormalization-group fixed points
$z=0$ and $z=\infty$, we have $\lambda=1$.
In the example presented above we have 
$\hat{r}_{g}(z_4,4)/\hat{r}_{g}(z_4,n) \approx 1.590$, so that 
$\lambda \approx 0.988$. The correction is small but not 
negligible.

As we discussed in Sec.~\ref{sec5.1}, a naive application of the 
transferability ideas would predict $z_n = z_4 (n/4)^{1/2}$. 
We can check how this approximation works in the present case.
If we set $z_{10} = z_4 (10/4)^{1/2}$, for our 
five values of $z_4$ we would obtain 
$z_{10} = 0.0889, 0.235, 0.506, 1.15, 3.97$, which are close to 
the estimates reported in Table \ref{table:transferability}.
 
\subsection{Comparison with full-monomer results}\label{sec5.3}

\begin{figure}[t]
\begin{center}
\begin{tabular}{c}
\epsfig{file=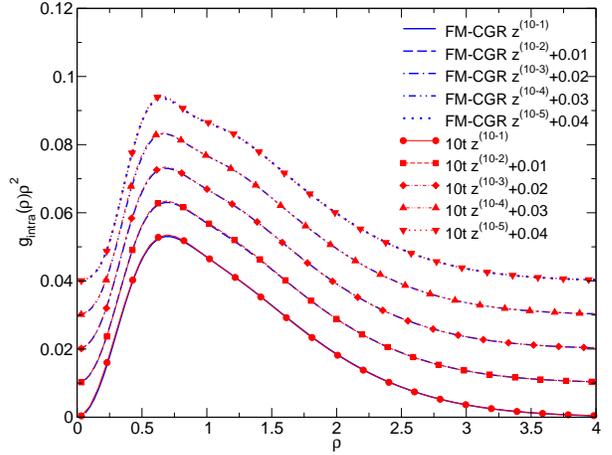,angle=0,width=8truecm} \hspace{0.5truecm} \\
\end{tabular}
\end{center}
\caption{Intramolecular distribution function $g_{\rm intra}(\rho)\rho^2$
at zero density as a function of $\rho=r/\hat{R}_g$.
We report decamer (10t) and full-monomer (FM-CGR) results for 
several values of $z$, see Table~\ref{table:transferability}. 
For the sake of clarity, results at different values of 
$z$ are shifted upward according to the legend.
}
\label{fig:gintra10}
\end{figure}

\begin{figure}[tbp]
\begin{center}
\begin{tabular}{c}
\epsfig{file=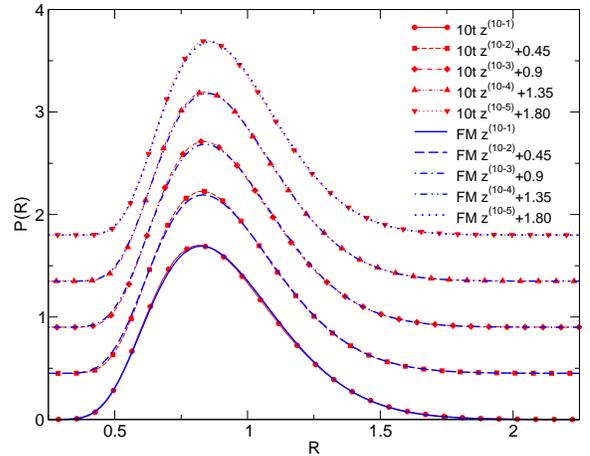,angle=0,width=8truecm} \hspace{0.5truecm} 
\end{tabular}
\end{center}
\caption{Distribution of the ratio $R=\hat{R}_{g,b}/\sqrt{\hat{R}^2_g}$ at 
zero density.
We report decamer (10t) and full-monomer (FM-CGR) results 
for several values of $z$, see Table~\ref{table:transferability}. 
For the sake of clarity, results at different values 
of $z$ are shifted upward according to the legend.
}
\label{fig:RG10}
\end{figure}

\begin{figure}[tbp]
\begin{center}
\begin{tabular}{c}
\epsfig{file=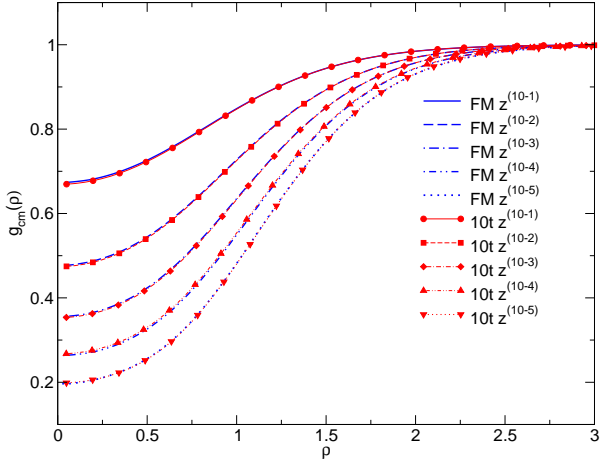,angle=0,width=8truecm} \hspace{0.5truecm} \\
\end{tabular}
\end{center}
\caption{Radial distribution functions for a pair of isolated chains 
as a function of  the center-of-mass separation $\rho = r/\hat{R_g}$. 
We report decamer (10t) and full-monomer (FM) results at zero density 
for several values of $z$, see Table~\ref{table:transferability}. 
}
\label{fig:GRCM10}
\end{figure}

By definition, our procedure is such that $A_2$ and the ratio 
$\hat{R}_{g,b}/\hat{R}_g$ at zero density are exactly reproduced. 
We wish now to check whether other structural and thermodynamic 
properties are satisfied.
In Figs.~\ref{fig:gintra10} and \ref{fig:RG10} 
we show the zero-density 
intramolecular distribution function $g_{\rm intra}(\rho)$
and the distribution of the CGR radius of gyration $\hat{R}_{g,b}$,
respectively. In both cases, the agreement is excellent, confirming that the
intramolecular structure is correctly reproduced. In Fig.~\ref{fig:GRCM10}
we show the zero-density center-of-mass 
pair distribution function for the decamer and 
for the polymer chain: 
again, the decamer appears to be quite accurate. Note that all
these results are far from obvious and indicate that our procedure has
correctly identified the renormalization-group flow from the 
$\theta$ point to the good-solvent fixed point.

\begin{table}[t]
\caption{For several values of $z$, see Table~\ref{table:transferability},
we report $A_3$ for polymers ($A_{3-FM}$),\protect\cite{CMP-08}
for the decamer ($A_{3-10}$), for 
the tetramer ($A_{3-4}$), and for the SB model $A_{3-SB}$.
}
\label{A3-decamer}
\begin {tabular}{ccccc}
\hline\hline
$z$ & $A_{3-FM}$ & $A_{3-10}$ & $A_{3-4}$ & $A_{3-SB}$ \\
\hline
$z^{(10-1)}$ & 0.216 & 0.217(1) & 0.2042(6) &  0.152  \\
$z^{(10-2)}$ & 1.18  & 1.191(5) & 1.132(4)  &  0.891  \\
$z^{(10-3)}$ & 2.76  & 2.78(1)  & 2.540(6)  &  2.160  \\
$z^{(10-4)}$ & 4.95  & 4.99(2)  & 4.91(1)   &  3.944  \\
$z^{(10-5)}$ & 7.96  & 7.95(3)  & 7.94(1)   &  6.396  \\
\hline\hline
\end{tabular}
\end{table}

Let us now consider the thermodynamic behavior. By construction, the 
decamer model reproduces the second-virial combination $A_2$. 
Let us now consider the third virial combination $A_3 = B_3 \hat{R}_g^{-6}$. 
In Table~\ref{A3-decamer} we report the results for the SB model,
the tetramer, and the decamer model, and compare them with 
the FM predictions.\cite{CMP-08} 
The SB model significantly underestimates 
$A_3$, as already observed in Sec.~\ref{sec4.1}. 
The tetramer model appears to be accurate 
close to the good-solvent limit, but some deviations are observed 
for intermediate values of $z$. The decamer model reproduces the FM results
with a relative accuracy of less than 1\% for all values of $z$, hence 
10 blobs are enough to reproduce quite accurately the 
thermodynamic behavior in the low-density regime.

\begin{table}[tbp]
\caption{Estimates of $K$ for the CGM with $n=4,10$ blobs 
and for the single-blob (SB) model. For the polymer model we report
the field-theory (FT) results of Ref.~\onlinecite{Schaefer-99},
see App.~\ref{App.B}. We also report the relative deviations 
$\Delta_n = 100 |K_n/K_{FT} - 1|$ with respect to the field-theory result.
Values of $z$ reported in Table~\ref{table:transferability}.
}
\label{tab:thermo-decamer}
\begin{tabular}{ccccccccc}
\hline \hline
$z$ & 
$\Phi$ & FT & $n=10$ & $n=4$  & SB & 
$\Delta_{10}\% $ & $\Delta_4\% $ & $\Delta_{SB}\% $\\
\hline
$z^{(10-5)}$ &
1& 4.45 &   4.42(2) & 4.29(1) & 3.91   & 0.7  & 3.6 & 12.1 \\
&2 & 8.98 &  8.64(7) & 8.135(25) & 7.02  & 3.8  & 9.5 & 21.8\\
&4 & 19.56   & 17.9(2) & 16.18(8) &   13.26  &  8.5  & 17.3 & 32.2 \\
&6 & 31.30 & 27.5(5) & 23.9(3) &	19.49 &  12.1 & 23.6 &	37.7 \\
\hline
$z^{(10-3)}$ &
1& 2.99 &   2.965(4) & 2.85(1) & 2.83   &  1.0 & 4.7 & 6.35 \\
&2 & 5.29  &  5.18(3) & 4.98(1) & 4.76  &  2.1 & 5.9 & 10.0\\
&4 & 10.23  & 10.00(6) & 9.19(4) & 8.64 &  2.2 &   10.2 & 15.5\\
&6 & 15.38& 14.6(1) & 13.8(2) & 12.52 & 5.0 & 10.3 & 18.6\\
\hline
$z^{(10-1)}$ &
1& 1.69 &   1.686(9) & 1.685(4) & 1.67 &  0.2 & 0.3 & 1.2 \\
&2 & 2.41  &  2.42(2) & 2.386(7) & 2.36  & 0.4 & 1.0 & 2 \\
&4 & 3.89  & 3.91(3) & 3.80(2) &   3.75  & 0.5 & 2.3 & 3.6\\
&6 & 5.39 & 5.33(8) & 5.28(7) & 5.15 & 1.1 & 1.9 & 4.7\\
\hline\hline
\end{tabular}
\end{table}

\begin{table}[tbp!]
\caption{RPA coefficients $k_{SB}(z)$ and $k_n(z)$, 
and corresponding full-monomer coefficient $k_{FM}(z)$; 
see Sec.~\ref{sec4.4} for the definitions. We also 
report the relative deviations
$\Delta_{10} = 100 |k_{10}(z)/k_{FM}(z) - 1|$ for the 
decamer.
Values of $z$ reported in Table~\ref{table:transferability}.
}
\label{tab:RPA10}
\begin{center}
\begin{tabular}{cccccc}
\hline
\hline
$z$ & $k_{SB}(z)$ & $k_4(z)$ & $k_{10}(z)$ & $k_{FM}(z)$ & $\Delta_{10}(z)\%$ \\
\hline
\hline
$z^{(10-1)}$ & 0.698 & 0.727& 0.751 & 0.798 & 6\%  \\
$z^{(10-2)}$ & 1.357 & 1.469& 1.563 & 1.798 & 13\%  \\
$z^{(10-3)}$ & 1.937 & 2.201& 2.397 & 3.042 & 21\% \\
$z^{(10-4)}$ & 2.502 & 2.976& 3.339 & 4.990 & 33\% \\
$z^{(10-5)}$ & 3.103 & 3.922& 4.587 & 10.39 & 56\% \\
\hline\hline
\end{tabular}
\end{center}
\end{table}

Finally, let us consider the finite-density behavior.  In the absence of 
FM simulations for $z^{(10-1)}$, $\ldots$, $z^{(10-5)}$, we cannot
directly compare structural properties. We will thus limit ourselves to compare 
the thermodynamic behavior, using the field-theory
expressions\cite{Schaefer-99} of App.~\ref{App.B}, to compute $K$ as a function
of $\Phi$. As we discussed in Sec.~\ref{sec4.2}, field theory appears
to be quite accurate, differences from the 
FM value being at most 4\%. In Table~\ref{tab:thermo-decamer}
we report $K(z,\Phi)$ for the different CGMs and compare 
it with the field-theory prediction. As expected, close to the $\theta$-point
($z = z^{(10-1)}$), the decamer reproduces very precisely the 
polymer result, at least for $\Phi\le 6$. To estimate the discrepancy 
for larger values of $\Phi$ [see
Eq.~(\ref{discrepancy-RPA})], we use the RPA estimates of the asymptotic
behavior reported in Table~\ref{tab:RPA10}. Discrepancies appear to be under
control for all values of $\Phi$, being at most 6\% in the limit
$\Phi\to\infty$. For $z=z^{(10-3)}$ differences are reasonable up to 
$\Phi\lesssim 6$. For larger densities the systematic deviations are larger
and $Z$ and $K$ are underestimated by 21\% for $\Phi\to\infty$. 
For $z=z^{(10-5)}$ the behavior is similar to that observed in the 
good-solvent regime:\cite{DPP-12-JCP} the decamer model is significantly 
more precise than the tetramer one and appears to be reliable up 
to $\Phi\approx 4$.

\section{Conclusions} \label{sec6}

Recently,\cite{DPP-12-Soft,DPP-12-JCP} we developed a consistent
coarse-graining strategy for polymer solutions under good-solvent conditions.
In this work we extend this strategy to the thermal crossover region. 
For large values of $L$, the universal features of the thermal crossover
can be completely characterized in terms of universal scaling functions,
which can be conveniently computed\cite{Sokal-94} 
by using the two-parameter model (TPM), 
at least not too close to the $\theta$ point.
Taking advantage of this relation, full-monomer results have been 
obtained by performing simulations of the 
Domb-Joyce model,\cite{DJ-72} which represents the lattice version of the 
TPM.  The TPM results are obtained by varying the on-site repulsion parameter 
$w$ together with the chain length $L$, in such a way to keep the 
product $wL^{1/2}$ fixed when taking the scaling limit $L\to\infty$. 
This procedure enables us to explore the thermal crossover region at fixed 
$z=(T-T_{\theta})L^{1/2}$ and reduced density $\Phi$ and, therefore, provides 
predictions for the scaling functions associated with any generic property
of the solution.

TPM zero-density scaling functions are used as target distributions to 
develop a 
CG model. As in our previous investigation for chains in the 
good-solvent regime, we have first developed a tetramer model, that is a 
model in which each chain is represented by four ``blobs.'' 
For the tetramer model we determine the interaction potentials at five
different values of the parameter $z$, that correspond to different 
solvent quality. The intramolecular potentials are defined in such a way 
to reproduce the structure of an isolated chain. For this purpose
we compute several single-chain scalar structural distributions
and then use the iterative Boltzmann inversion procedure to 
determine the intramolecular potentials. 
Intermolecular interactions are specified by a single blob-blob pair potential
which is determined by 
matching the radial distribution function 
between the centers of mass of the chains. The tetramer CG model set up at 
zero density is found to reproduce the collective behavior at 
finite density reasonably well up to a density $\tilde{\Phi}(z)$ 
which decreases with increasing $z$. 
For small values of $z$, i.e., close to the $\theta$
regime, the tetramer model can be safely used up to very high $\Phi$ 
(the error on $Z$ is at most 1\% up to $\Phi\sim 10$), as long as 
$L$ is large enough to avoid tricritical effects. On the other hand,
close to the good-solvent regime, 
we recover our previous finding that the tetramer model is reliable up to 
$\tilde{\Phi}\sim 2$.

In order to enlarge the range of applicability of the CG model with density, 
we have also developed a transferability procedure which allows us to use 
the tetramer potentials to build CG models with more blobs per chain. 
While in the good-solvent regime a simple rescaling of the characteristic 
length scale of the tetramer potentials was found to be sufficient to provide 
models with higher CG resolutions, in the crossover regime 
we must both change the basic length scale and temperature, i.e., $z$, 
as the number of blobs is increased.
Using this more elaborate procedure,
 we have transferred the tetramer potentials to a CG model with 10 blobs 
per chain. As expected, 
the density range in which the predictions of the decamer CG model are 
accurate is enlarged with respect to the tetramer case.

This work completes our effort to  develop a CG strategy for polymer solutions
which employs potentials derived at zero density, but which is still able 
to predict the correct thermodynamics and structural properties of the system 
at finite density, deep into the semidilute regime. Since the potentials 
are derived at zero density, we avoid all inconsistencies related to the 
use of state-dependent potentials,\cite{SST-02,Louis-02,DPP-13-JCP}
which plague most of the CG models employed to study complex fluids. 
Our CG  model can
be used to investigate temperature effects in polymer solutions,
hence to compare with experimental results obtained by using chains of 
limited extension and that lie in this intermediate region of the phase 
diagram. 
Extensions of our strategy to treat polymer-wall and polymer-colloid 
interactions, polymers of different architecture,  and copolymers 
are under investigation and will open the way to a fully consistent CG 
modeling of more challenging and interesting systems such as polymer solutions 
of various solvent quality in the presence of an absorbing wall 
(depletion interactions), colloid-polymer solutions,
 and block copolymer solutions, the behavior of which is well characterized
 by experiments but not so well by theory. 

\appendix
\section{The blob radius of gyration} \label{App.A}

\begin{table}[h]
\caption{Estimates of the ratios $\hat{R}_{g,b}^2/\hat{R}_g^2$ 
for $n=4$ and $n = 10$ blobs for several values of $z$, see Table~\ref{tab-z}. }
\label{tab:rgratios}
\begin{tabular}{ccc}
\hline\hline
 $z$ & $n=4$ &  $n=10$ \\
 \hline
 $ 0$	 &   0.7500    & 0.9000 \\
 $z^{(1)}$ & 0.7553(3) & 0.9032(2) \\
 $z^{(2)}$ & 0.7612(2) & 0.9068(3) \\
 $z^{(3)}$ & 0.7686(5) & 0.9114(2) \\
 $z^{(4)}$ & 0.7769(5) & 0.9167(2) \\
 $z^{(5)}$ & 0.7874(4) & 0.9236(1) \\
 $\infty$&   0.7959(2) & 0.9295(2)\\
\hline\hline
\end{tabular}
\end{table}
In this Appendix we wish to compute the function 
$S_b(A_2,n)=\hat{R}_{g,b}(z,n)/\hat{R}_g(z)$ for $n=4$ and 
$n=10$, which we parametrize in terms of $A_2$ instead of $z$. 
We first determine its behavior for $A_2\to 0$, by 
performing a one-loop computation in the two-parameter model.
\cite{Schaefer-99,dCJ-book}
We begin by considering the average quadratic distance
between monomers $i$ and $j$. If $L$ is the total number of monomers of the
chain, i.e. its contour length, $x = i/L$ and $y = j/L$, we have for 
$j > i$:
\begin{eqnarray}
{1\over \hat{R}^2_{g,0}} \langle ({\bf r}_i - {\bf r}_j)^2 \rangle &=&
   6(y-x) + {8\over3} z 
   \left[ 12 x y^{1/2} - 8 x^{3/2}\right.
\\ 
&& \left.
    - 4 y^{3/2} +
    8 (y-x)^{3/2} + 3 (x-y)^2 \right. 
\nonumber \\
&& \left. + 4 (2 + x - 3 y) \sqrt{1-x} - 
    8 (1 - y)^{3/2} \right],
\nonumber 
\end{eqnarray}
where $\hat{R}^2_{g,0}$ is the radius of gyration for the ideal case ($z=0$). 
We define the swelling factor associated with  $\hat{r}_g$ as 
\begin{equation}
 \alpha_{g} (n,z) = {\hat{r}^2_g(n,z) \over \hat{r}^2_g(n,0)}.
\end{equation}
Using the expression reported above we obtain to first order in $z$:
\begin{eqnarray}
\alpha_g(1,z) &=& 1 + {134\over 105} z, \\
\alpha_g(2,z) &=& 1 + {1\over 105} (912 \sqrt{2} - 1181) z 
\nonumber \\ &= &
 1 + 1.036 z, \\
\alpha_g(3,z) &=& 1 + 0.914 z, \\
\alpha_g(4,z) &=& 1 + 0.833 z. 
\end{eqnarray}
For $n\ge 5$, the large-$n$ approximation
\begin{equation}
\alpha_g(n,z) = 1 + z \left({256\over 105} {1\over \sqrt{n}} - {2\over n} + 
     {0.98\over n^{3/2}} \right),
\end{equation}
works well. Since 
\begin{equation}
S_b(A_2,n) = {\hat{R}_{g,b}^2\over \hat{R}_g^2} = 
  1 - {\hat{r}_g^2\over \hat{R}_g^2} = 
  1 - {\hat{r}_{g,0}^2\over \hat{R}_{g,0}^2} {\alpha_g(n,z)\over
       \alpha_g(1,z)},
\end{equation}
We obtain for $z\to 0$ 
\begin{eqnarray}
&& S_b(A_2,4) = 0.75 + 0.110793 z = 0.75 + 0.00497423 A_2,
    \nonumber \\
&& S_b(A_2,10) = 0.90 + 0.067556 z = 0.90 + 0.00303304 A_2,
\nonumber \\
\end{eqnarray}
where we used $z = 2 A_2 (4 \pi)^{-3/2}$ to leading order. 
To obtain the behavior for all values of $A_2$, we use the 
results reported in Table~\ref{tab:rgratios}, which have been obtained 
by FM simulations of the DJ model. Interpolating the data we obtain
\begin{eqnarray}
S_b(A_2,4) &=& \left(0.75 + 0.00497423 A_2 +0.00023944 A_2^2 
  \right. \nonumber \\
&&\left.
    +0.0000679109 A_2^3\right)^{1/2}, \\
S_b(A_2,10) &=& \left(0.9 + 0.00303304 A_2 + 0.000509387 A_2^2 \right.
    \nonumber \\ 
    && \left. - 
           0.000153285 A_2^3  + 
           0.0000250389 A_2^4 \right)^{1/2}.
\nonumber 
\end{eqnarray}
Given $S_b(n)$, we can then compute $F_{A,b}(n) = A_2/S_b(n)^3$.

\section{Field-theory predictions} \label{App.B}

We summarize here the field-theoretical results of Sch\"afer,
\cite{Schaefer-99} reporting 
the basic formulae which allow one to compute the compressibility
factor in terms of $\Phi$ and $z$. In this approach the crossover and 
density behavior is parametrized by two independent variables $f$ and $w$.
The variable $f$ parametrizes the crossover from ideal ($f=0$) to 
good-solvent behavior ($f=1$), while $w$ parametrizes the density dependence,
$w=1$ corresponding to $\Phi=0$ and $w = 0$ to $\Phi=\infty$.
To relate them to $z$ and the polymer volume fraction $\Phi$, we need to 
define several auxiliary functions (see Chap. 13 of
Ref.~\onlinecite{Schaefer-99}):
\begin{eqnarray}
H(f) &=& 1 - 0.005 f - 0.028 f^2 + 0.022 f^3, \\
H_u(f) &=& (1 + 0.824 f)^{0.25}, \\
H_n(f) &=& H(f) H_u(f)^{-2}.
\end{eqnarray}
Then, we define (Eqs. (13.27) and (15.12) of Ref.~\onlinecite{Schaefer-99}):
\begin{eqnarray}
\alpha^2_g(f) &=& (1-f)^{(1-2\nu)/\nu\omega} H(f) (1 - 0.195 f), \\
\tilde{z}(f,w) &=& f (1-f)^{-1/(2\nu\omega)} H_n(f)^{-1/2} \sqrt{n_0}/w, 
\\
\tilde{s}(f,w) &=& {c_0 n_0\over \tilde{u} \tilde{z}(f,w)} 
   (1-f)^{(3\nu-2)/\omega\nu} \nonumber \\
&& \times H_u(f) H(f)^{-2} (1-w^2)/w^2,
\end{eqnarray}
where $\tilde{u} = 8.1075$, $c_0 = 1.2$, $n_0 = 0.53$, $\nu = 0.588$,
$\omega = 0.80$. Here $\alpha_g$ is the usual expansion factor,
$\alpha_g =\hat{R}_g/\hat{R}_{g0}$, where $\hat{R}_{g0}$ is the 
value of the radius of gyration for the ideal chain,
$\tilde{z}$ corresponds to the crossover variable $z$ apart 
from a normalization ($z = 0.182 \tilde{z}$), while $\tilde{s}$ is 
$c N^{3/2}$, where $c$ is the concentration.

Given $z$ and $\Phi$, to obtain $f$ and $w$ we work as follows. 
First, we compute $\tilde{f}$ such that 
$\tilde{z}(\tilde{f},1)=z/0.182$ ($\tilde{f}$ and $w=1$ correspond to
$z$ and $\Phi=0$ in our variables) and 
\begin{equation}
   s_t = {3 \Phi\over 4 \pi} \alpha_g(\tilde{f})^{-3}.
\end{equation}
Then, we determine $f$ and $w$ by solving the equations
\begin{equation}
    \tilde{z}(f,w) = z/0.182 \qquad\qquad
    \tilde{s}(f,w) = s_t.
\end{equation}
The compressibility factor is then determined as 
(Eqs. (17.21) and (17.51) of Ref.~\onlinecite{Schaefer-99}):
\begin{eqnarray}
Z(f,w) &=& 1 + {1\over 2} (1 + 2 u^* f) W_R 
 \nonumber \\
  && - {f u^* \sqrt{\pi N_R}\over 3 W_R} 
       {0.808 + 1.22 W_R \over 1 + 1.22 W_R} 
\\
  && \times   \left[1 + (W_R - 1) (1 + 2 W_R)^{1/2} \right],
\nonumber 
\end{eqnarray}
where $u^* = 0.364$, $W_R$ and $N_R$ are functions of $w$ given by
\begin{equation}
W_R = c_0 n_0 (1 - w^2)/w^2 \qquad\qquad
N_R = n_0/w^2.
\end{equation}


\begin{thebibliography}{99}
\bibitem{footPsi}
In experimental work the low-density behavior of
the osmotic pressure $\Pi$ is usually written as
$\Pi/(R T\rho) = 1/M + B_{2,\rm expt} \rho + O(\rho^2)$, where
$M$ is
the molar mass of the polymer, $\rho$ the weight concentration, and
$T$ the absolute temperature. Then,
$B_2 \equiv  M^2 B_{2,\rm expt}/N_A$, where $N_A$ is the
Avogadro number.
\end{thebibliography}
\end{document}